\documentclass[conference]{IEEEtran}
\IEEEoverridecommandlockouts

\PassOptionsToPackage{table,xcdraw}{xcolor}

\usepackage{cite}
\usepackage{amsmath,amssymb,amsfonts}
\usepackage{graphicx}
\usepackage{textcomp}
\usepackage{xspace}
\usepackage{ifthen}
\usepackage[inline]{enumitem}
\usepackage{subcaption}
\usepackage[hidelinks]{hyperref}
\usepackage{cleveref} 
\usepackage{todonotes}
\usepackage{csquotes}
\usepackage{url}
\usepackage{units}
\usepackage{tikz}
\usepackage{etoolbox}
\usepackage{caption}
\usepackage{verbatim}
\usepackage{pgfplots}
\usepackage{booktabs}
\usepackage{pdfpages}
\usepackage{fancyhdr}
\usepackage{framed}
\usepackage[printonlyused]{acronym}
\usepackage{relsize}
\usepackage{balance}
\usepackage{microtype}
\usepackage{censor}
\usepackage[normalem]{ulem}

\usepackage{rotating} 
\usepackage{multirow}
\usepackage{xcolor}
\usepackage{tabularx}

\usepackage{pgfplots}
\pgfplotsset{width=\columnwidth,compat=1.9}

\crefformat{section}{\S#2#1#3}
\crefformat{subsection}{\S#2#1#3}
\crefformat{subsubsection}{\S#2#1#3}
\crefmultiformat{section}{\S#2#1#3}{ and~\S#2#1#3}{, \S#2#1#3}{, and~\S#2#1#3}
\crefformat{figure}{Fig.\nobreak\hspace{0.25em}#2#1#3}
\crefformat{algorithm}{Alg.\nobreak\hspace{0.25em}#2#1#3}
\crefformat{table}{Tab.\nobreak\hspace{0.25em}#2#1#3}
\crefmultiformat{figure}{Figure #2#1#3}{ and~Figure #2#1#3}{, Figure #2#1#3}{, and~Figure#2#1#3}

\usepackage{pgfplots}
\pgfplotsset{width=\columnwidth,compat=1.9}
\usetikzlibrary{patterns}

\usetikzlibrary{external, positioning, shapes, arrows, fit, backgrounds, calc, matrix}

\newboolean{showcomments}
\setboolean{showcomments}{true}
\ifthenelse{\boolean{showcomments}}
{ \newcommand{\mynote}[3]{
		\textcolor{#3}{{\bfseries\sffamily\scriptsize#1: }\small#2}
}}
{ \newcommand{\mynote}[3]{}}

\newcommand{\ie}{i.\,e.,\xspace}
\newcommand{\eg}{e.\,g.,\xspace}

\newcommand{\cf}{c.\,f.,\xspace}

\def\BibTeX{{\rm B\kern-.05em{\sc i\kern-.025em b}\kern-.08em
    T\kern-.1667em\lower.7ex\hbox{E}\kern-.125emX}}

\usepackage[absolute,showboxes]{textpos}
\TPMargin{5pt}
\textblockcolour{gray!02}

\newcommand{\copyrightstatement}{
	\begin{textblock}{1}(0.00,0.00)    
		\noindent
		\scriptsize
		\copyright \  
		2023 IEEE.
		Personal use of this material is permitted.  Permission from IEEE must be obtained for all other uses, in any current or future media, including reprinting/republishing this material for advertising or promotional purposes, creating new collective works, for resale or redistribution to servers or lists, or reuse of any copyrighted component of this work in other works.
		This paper is the authors' accepted version to be published in the 5th IEEE International Conference on Blockchain and Cryptocurrency (ICBC). For the final, published version we refer to DOI [\textit{to be inserted here later upon publication}].
	\end{textblock}
}

\begin{document}
\copyrightstatement
\title{SoK: Scalability Techniques for BFT Consensus}

\author{\IEEEauthorblockN{Christian Berger\IEEEauthorrefmark{1}\IEEEauthorrefmark{2},
Signe Schwarz-Rüsch\IEEEauthorrefmark{1}\IEEEauthorrefmark{3},
Arne Vogel\IEEEauthorrefmark{1}\IEEEauthorrefmark{3}, \\
Kai Bleeke\IEEEauthorrefmark{4},
Leander Jehl\IEEEauthorrefmark{5},
Hans P. Reiser\IEEEauthorrefmark{6},
and Rüdiger Kapitza\IEEEauthorrefmark{3}} \\
\IEEEauthorblockA{\IEEEauthorrefmark{1}
The first three authors contributed equally to this work}
\IEEEauthorblockA{\IEEEauthorrefmark{2}University of Passau,
Passau, Germany,
Email: cb@sec.uni-passau.de}
\IEEEauthorblockA{\IEEEauthorrefmark{3}
Friedrich-Alexander-Universtität Erlangen-Nürnberg,
Erlangen, Germany,
Email: \{ruesch, vogel, kapitza\}@cs.fau.de}
\IEEEauthorblockA{\IEEEauthorrefmark{4}
Technische Universität Braunschweig,
Braunschweig, Germany, Email: bleeke@ibr.cs.tu-bs.de}
\IEEEauthorblockA{\IEEEauthorrefmark{5}
University of Stavanger,
Stavanger, Norway,
Email: leander.jehl@uis.no}
\IEEEauthorblockA{\IEEEauthorrefmark{6}
Reykjavik University,
Reykjavik, Iceland,
Email: hansr@ru.is}
}

\IEEEoverridecommandlockouts
\IEEEpubid{\makebox[\columnwidth]{978-8-3503-1019-1/23/\$31.00~\copyright2023 IEEE \hfill} \hspace{\columnsep}\makebox[\columnwidth]{ }}
 
\maketitle
\IEEEpubidadjcol

\acrodef{BFT}{Byzantine Fault Tolerant}
\acrodef{SGX}{Intel's Software Guard Extensions}
\acrodef{PoW}{Proof-of-Work}
\acrodef{PoS}{Proof-of-Stake}
\acrodef{HLF}{Hyperledger Fabric}
\acrodef{TEE}{trusted execution environment}
\acrodef{SCM}{supply chain management}
\acrodef{DLT}{distributed ledger technology}
\acrodef{CFT}{crash-fault tolerant}
\acrodef{PBFT}{Practical Byzantine Fault Tolerance}
\acrodef{PoET}{Proof-of-Elapsed-Time}
\acrodef{TPM}{Trusted Platform Module}
\acrodef{SEV}{Secure Encrypted Virtualization}
\acrodef{SNP}{Secure Nested Paging}
\acrodef{UTXO}{Unspent Transaction Output}
\acrodef{DeFi}{decentralized finance}
\acrodef{VRF}{verifiable random function}
\acrodef{SMR}{State machine replication}
\acrodef{PoE}{Proof-of-Execution}
\begin{abstract}
With the advancement of blockchain systems, many recent research works have proposed
distributed ledger technology~(DLT) that employs Byzantine fault-tolerant~(BFT) consensus
protocols to decide which block to append next to the ledger. Notably, BFT consensus 
can offer high performance, energy efficiency, and provable correctness properties, and it is thus
considered a promising building block for creating highly resilient and performant blockchain infrastructures.
Yet, a major ongoing challenge is to make BFT consensus applicable to large-scale environments.
A large body of recent work addresses this challenge by developing novel ideas to improve the scalability of BFT consensus,
thus opening the path for a new generation of BFT protocols tailored to the needs of blockchain.
In this survey, we create a systematization of knowledge about the novel scalability-enhancing techniques that state-of-the-art BFT consensus protocols use. For our comparison, 
we closely analyze the efforts, assumptions, and trade-offs these protocols make.
\end{abstract}


\section{Introduction}
\label{sect:introduction}

Blockchain-based \ac{DLT} experienced increasing adaptation and growing popularity in recent years.
The original Bitcoin protocol uses a \ac{PoW} scheme to achieve agreement on blocks of transactions, also called Nakamoto consensus~\cite{nakamoto2008bitcoin}.
Bitcoin and many of its more recently upcoming competitors aim to realize secure and decentralized applications. 
While Bitcoin's application is to allow its users to send and receive digital peer-to-peer cash, other blockchains, such as Ethereum~\cite{buterin2014next}, even allow building more complex applications by submitting generic code (called smart contracts) to the blockchain, where functions of smart contracts can be invoked by users sending transactions. 
A blockchain's security and decentralization depend on a large number of network participants. 
At the same time, the system performance, \ie the throughput, should suffice to match the application's requirements. 

\paragraph{Agreement for Blockchain Networks} Blockchains use an \textit{agreement protocol} to decide which block they append next to the ledger. 
Ideally, agreement should work on a large-scale (meaning \emph{many} nodes) and geographically dispersed environment. 
\ac{PoW} achieved agreement and successfully allowed open membership by securing the blockchain network against Sybil attacks~\cite{douceur2002sybil}. 
This is because \ac{PoW} couples the probability of a node being allowed to decide the next block towards the computational resources it utilized over a certain time span. 
Because \ac{PoW} achieves agreement \textit{without coordination between the nodes} other than disseminating the decided blocks, it can scale well for a large number of nodes. 
As of this writing,  there are over 15,000 reachable Bitcoin nodes\footnote{https://bitnodes.io/}.

Nevertheless, \ac{PoW} comes with inherent design problems. 
In particular, it (1) wastes energy and computing resources, (2) usually does not scale up its performance when utilizing more resources, thus making the scheme very inefficient, and (3) it does not guarantee consensus finality~\cite{vukolic2015quest}, a property that ensures a block, once decided, is never changed later on.

\paragraph{Coordination-based \ac{BFT} Agreement}
Recent research papers try to work around these problems by proposing to utilize \textit{coordination-based BFT agreement protocols} (which we will simply refer to as BFT protocols as of now) used in well-conceived protocols like PBFT~\cite{castro1999practical}, initially proposed more than 20 years ago. 
The benefits of PBFT and related protocols like BFT-SMaRt~\cite{bessani2014state} lie in proven protocol properties and various performance optimizations allowing these protocols to achieve up to the magnitude of $10^5$ transactions per second. 

In particular, BFT algorithms can be used as a \ac{PoS} variant~\cite{buterin2017casper}, in which blockchain nodes are granted permission to participate in the agreement, depending on the stake (\ie native cryptocurrency of the blockchain) they own. 
In coordination-based BFT protocols, the decision about which block is being appended to the blockchain is canonical among all correct nodes~\cite{buterin2017casper}; thus, they ensure consensus finality, and they are at the same time energy-efficient, \ie energy is only consumed for meaningful computations.


\subsection{Motivation}

\begin{figure}[t]
    \centering
      \begin{tikzpicture}
    \begin{axis}[
width=7.2cm,
height=4cm,
font= \footnotesize, 
    xlabel={Replicas},
    ylabel={Throughput [kOps/s]},
    xmin=0, xmax=128,
    ymin=0, ymax=200,
    xtick={ 4, 16, 32, 64, 128},
    ytick={0, 50, 100,150, 200},
    legend pos=south east,
    legend columns = 1,
    legend style={at={(0.98, 0.6)}},
    legend cell align={left},
    ymajorgrids=true,
    xmajorgrids=true,
    grid style=dashed,
]
 
\addplot[
    color=blue,
    mark=square,
    ]
    table [x=replicas,y=throughput] {data/bftsmart-p128.txt};
\addplot[
    color=red,
    mark=triangle,
    ]
    table [x=replicas,y=throughput] {data/bftsmart-p1024.txt};
   
  \legend{\footnotesize 128 byte payload, 1024 byte payload}

\end{axis}
\end{tikzpicture} 
    \caption{Performance of BFT-SMaRt (measured by~\cite{yin2018hotstuff}).}
    \label{fig:intro:bftsmart-performance}
\end{figure}
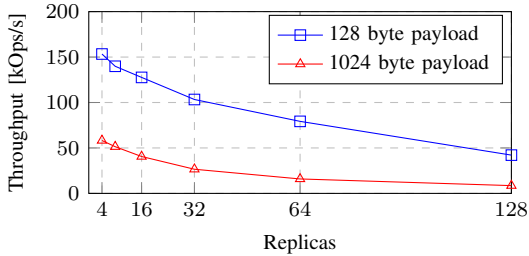

\begin{figure}[t]
    \includegraphics[width=\columnwidth]{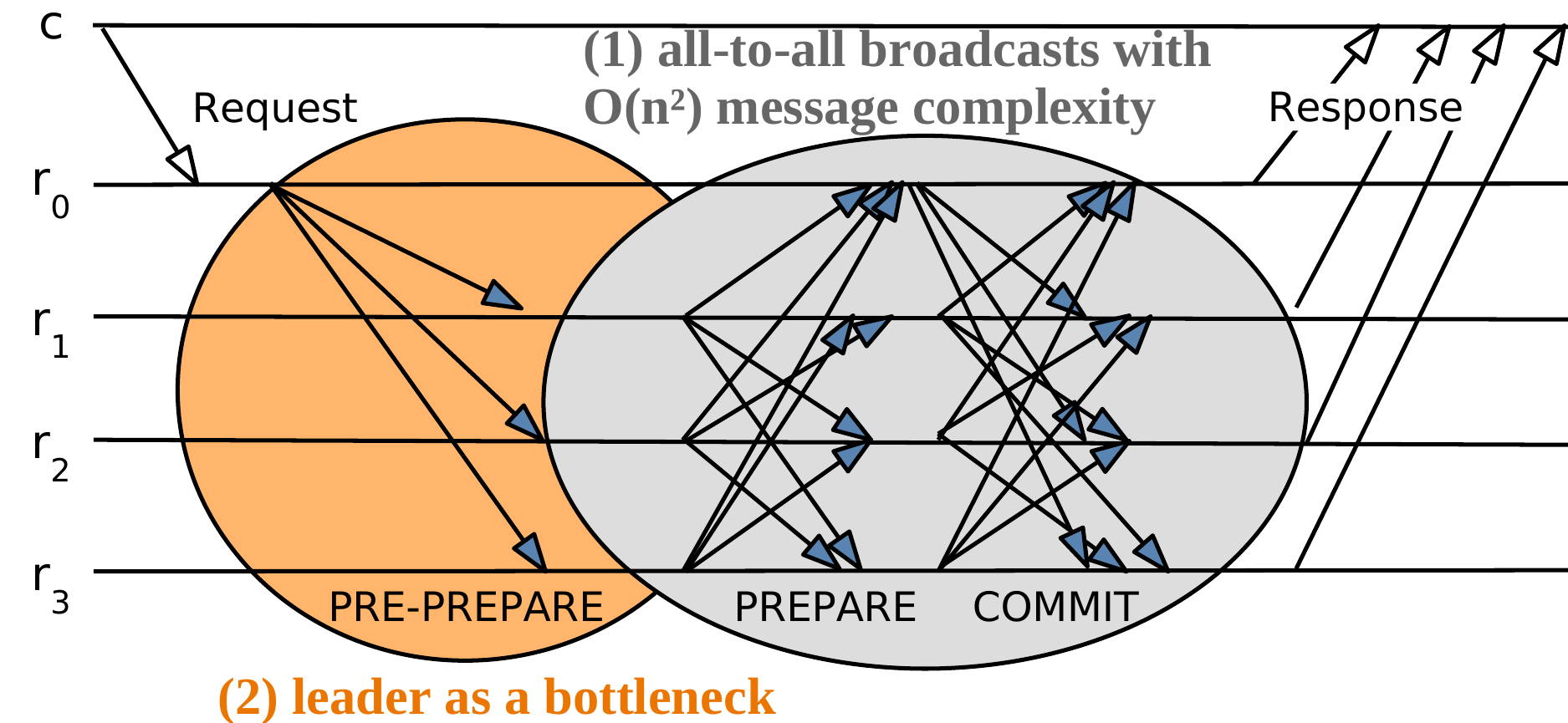}
    \caption{Scalability problems of PBFT and BFT-SMaRt.}
    \label{fig:intro:pbft-scalability}
\end{figure}

\paragraph{The Scalability Challenge for BFT Agreement}
Traditional BFT protocols have a problem with scaling to large system sizes. 
To illustrate this with a brief example, let us review the well-known PBFT protocol, which is efficient (with performance in the magnitude of $10^5$ transactions per second) for small-sized systems but suffers a noticeable performance decrease (of down to a magnitude of $10^3$) when several hundred nodes are in the system. 
\cref{fig:intro:bftsmart-performance} shows the declining performance of a newer protocol, BFT-SMaRt (with the same agreement pattern as PBFT), when the system size increases.

PBFT and BFT-SMaRt share two main problems (see \cref{fig:intro:pbft-scalability}): First, the \textit{normal operation} message complexity is $O(n^2)$ because all replicas exchange their votes using all-to-all broadcasts. 
This makes the underlying communication topology of the protocol  essentially a clique.
Further, the more nodes are in the system, the more resources are consumed for only verifying message authenticators from all other nodes. 

Second, the protocol flow seems imbalanced: An additional burden is being put on the leader because the leader needs to receive transactions from all clients and then disseminate them in a batch to all other replicas. 
This essentially makes the leader's bandwidth or its computational capabilities to produce message authenticators a limiting factor for the performance of the BFT protocol since all transactions of the system need to be channeled through a single leader.

\paragraph{The Age of Novel BFT Blockchains}
Research on BFT consensus becomes increasingly necessary and practical.  Recently, many BFT protocols have been proposed for usage in blockchain infrastructures, such as HotStuff~\cite{hotstuff19}, SBFT~\cite{gueta2019sbft}, Tendermint~\cite{buchman2016tendermint, cason2021design},  Algorand~\cite{gilad2017algorand}, Avalanche~\cite{rocket2020avalanche}, Mir-BFT~\cite{stathakopoulou2019mir}, RedBellyBC~\cite{crain2021red}, and Kauri~\cite{neiheiser2021kauri}. 

These protocols aim at making BFT consensus more scalable, thus delivering high throughput at low latency in systems with hundreds or even thousands of participants. Scalability is essential to allow a blockchain to grow its ecosystem and satisfy the demands of many \ac{DeFi} applications.

For instance, as of the time of writing, the Avalanche mainnet consists of 1294 validators\footnote{https://explorer-xp.avax.network/validators} and has a total value locked of 2.78 billion USD\footnote{https://defillama.com/chain/Avalanche}.
But which techniques does the blockchain generation of BFT protocols employ to improve its scalability over well-known BFT protocols such as PBFT? In this survey, we strive to explore the vast design space of novel protocols and analyze their ideas for scaling Byzantine consensus.

\paragraph{A Closer Look on Scalability-Enhancing Techniques}
As shown in \cref{fig:intro:pbft-scalability}, traditional \ac{BFT} protocols do not scale well.  
One way to improve the scalability of BFT protocols is to optimize the protocol logic in a way that (1) reduces bottleneck situations and distributes the transaction load as evenly as possible among the available capacities of every replica and (2) utilizes clever aggregation techniques to reduce the overall message (or: authenticator) complexity in the system.

Many more abstract ideas have been developed to make consensus more scalable --  these ideas concern, for example, optimizing communication flow, parallelizing consensuses (\ie \textit{sharding}), utilizing special cryptographic primitives, or using trusted hardware components.

\subsection{Research Questions}

The newer \enquote{blockchain generation} of BFT protocols requires scalable Byzantine consensus, which made exploring, advancing, and combining several scalability-enhancing techniques a broad and ongoing research field. In this paper, we want to create a systematization of knowledge on these ongoing efforts, which leads to our two main research questions:

 \begin{itemize}
\item[\textbf{R1}] Which novel techniques exist for scaling Byzantine consensus?
\item[\textbf{R2}] How do recent research papers combine existing scalability-enhancing techniques or ideas in a novel way to achieve better scalability than traditional BFT protocols?
 \end{itemize}

Scalability here relates to the number of nodes in the system.
We analyze which techniques can improve scalability in BFT protocols, \eg compared to traditional protocols like PBFT, and also cover the selection of smaller committees as a scalability technique.
Surveying the mechanisms of how to provide open membership is not the focus of this work, and we do not focus on defenses against Sybils in open membership systems, \eg computing a proof of work or depositing stake.

\subsection{Contributions}

The main ambitions of our survey paper are to review state-of-the-art research papers on BFT protocols to identify and classify the scalability-enhancing techniques that have been developed.
We further investigate the assumptions, ambitions, and trade-offs with which these techniques are used. In particular, our main contributions are the following:
\begin{itemize}[leftmargin=1em, itemsep=.1em, parsep=.1em, topsep=.1em,partopsep=.1em]
	\item We conduct a systematic search for exploring scalability-enhancing techniques for \ac{BFT} consensus.
	\item Moreover, we create a taxonomy that classifies and summarizes all of the found techniques. 
	\item Further, we also create a comparison of the new generation of scalable \ac{BFT} protocol designs.
	\item We comprehensively discuss the different ideas on an abstract level and pinpoint the design space from which these \ac{BFT} protocols originate.
\end{itemize}

\subsection{Outline}
In \cref{background}, we first give an overview of the basics of \ac{BFT} protocols. 
Next, we present our methodology in \cref{sect:methodology}, which is based on a systematic literature search. 
Further, in \cref{sect:scalability}, we present our survey on scalability-enhancing techniques for Byzantine consensus, trying to answer the research questions from above. 
After that, we summarize the efforts of related surveys in \cref{sect:related_work} and conclude in \cref{sect:conclusions}.


\section{BFT in a nutshell}
\label{background}



In this section, we review a few basics of \ac{BFT} protocols.

\subsection{Assumptions}

\subsubsection{The Byzantine Fault Model}

A faulty process may behave arbitrarily in the Byzantine fault model, even exhibiting malicious and colluding behavior with other Byzantine processes. The model always assumes that only a threshold $f$ out of $n$ participants is Byzantine, while all others ($n-f$) are correct and show behavior that exactly matches the protocol description. 
Although described as arbitrary, the behavior and possibilities of a Byzantine process are still limited by its resources and computational feasibility, e.g., it can not break strong cryptographic primitives. Lamport showed with the
Byzantine generals' problem that achieving consensus with $f$ Byzantine participants is impossible 
if $f \geq n/3$ and solvable for $f < n/3 $ in the partially synchronous or asynchronous model (\cf \cref{sec:background:synchrony})~\cite{pease1980reaching, lamport1982byzantine, dwork1988consensus}.
An adversary is often modeled as being in control over all $f$ Byzantine processes. The adversary's access to information can be \textit{limited} through private channels between replicas or \textit{unlimited} if the adversary is modeled to have full disclosure on all messages sent over the network. Lastly, the adversary is assumed either \textit{static}, i.e., has to make his selection initially without the possibility to change it later on, or \textit{adaptive}, in which the adversary can change the nodes as long as not exceeding the threshold $f$ at any given time.


\subsubsection{Synchrony Models}\label{sec:background:synchrony}

BFT protocols rely on synchrony models to capture temporal behavior and timing assumptions, which are important for the concrete protocol designs. 
In this subsection, we review popular models (see \cref{table:background:synchrony}) and discuss how they affect  practical system design.

\paragraph{Asynchronous System Model}
No assumptions are made about upper bounds for network transmissions or performing local computations. These are said to complete \textit{eventually}, meaning they happen after an unknown (but \textit{finite}) amount of time. For instance, the network cannot \enquote{swallow up} messages by infinitely delaying them. The asynchronous model is the most general, yet it complicates the design of protocols in this model since no timers can be used in the protocol description. It was proven impossible to deterministically achieve consensus in the presence of faults in an asynchronous system~\cite{fischer1985impossibility}. This problem is solvable in a synchronous system\cite{pease1980reaching} where timers can be used to detect failures.

\paragraph{Synchronous System Model}
Here we assume that strict assumptions can be made about the timeliness of all events. In particular, the synchronous system model assumes the existence of a known upper bound $\delta$ for the time needed for both message transmissions over the network and local computations. In practice, choosing $\delta$ to model a system can be bothersome: If the correctness of decisions depends on it, it must not be underestimated; however, if it is chosen too large, it may negatively impact the performance of a system.

\paragraph{Partially Synchronous System Model}
Dwork et al. proposed a sweet spot between the synchronous and asynchronous model~\cite{DLS88}. Partial synchrony comes in two versions: \textit{Unknown bounds} partial synchrony assumes an unknown bound that always holds. \textit{Global stabilization time (GST)} partial synchrony assumes the bound is initially known but only holds \textit{eventually}, \ie after some unknown time span which is modeled by the GST. The latter is often also referred to as \textit{eventual synchrony} since the system behaves exactly like a synchronous system as soon as GST is reached. Partial synchrony is popular among many BFT protocols (like PBFT) that guarantee liveness only under partial synchrony but always remain safe even when the system is asynchronous.

\begin{table}[!tbp]
	\centering
	\resizebox{\columnwidth}{!}{ 
	\begin{tabular}{*3c}
		\toprule
		System model & Bound $\delta$ exists?  & Bound holds ...\\
		\midrule

		Synchronous & known $\delta$ & always  \\

		Eventually synchronous & known $\delta$ & after unknown GST    \\
		Partially synchronous  & unknown $\delta$ & always  \\

		Asynchronous   & \multicolumn{2}{c}{unbounded } \\
		\bottomrule
	\end{tabular}
	}
\caption{Overview over different synchrony models.}
\label{table:background:synchrony}
\end{table}

\subsection{State Machine Replication and Consensus}

\textit{\ac{SMR}} is a technique for achieving fault tolerance by replicating a centralized service on several independent replicas, which emulate the service. 
Replicas agree on inputs (transactions) proposed by clients and thus ensure a consistent state and matching results. 
Such agreement is typically achieved through a consensus protocol.
A \textit{consensus} protocol guarantees that all correct participants eventually decide on the same value from a set of proposed input values. Moreover, an SMR protocol additionally requires \textit{output consolidation}: The client needs to collect at least $f+1$ matching responses from different replicas to assert its correctness. Formally, SMR should satisfy the following guarantees:

\begin{itemize}[leftmargin=1em, itemsep=.1em, parsep=.1em, topsep=.1em,partopsep=.1em]

\item \textit{Safety} (\textit{Linearizability}): The replicated service 
behaves like a centralized implementation that executes transactions atomically one at a time~\cite{castro2000phd}.

\item \textit{Liveness} (\textit{Termination}): Any transaction issued by a correct client eventually  completes~\cite{lynch1996distributed}.
\end{itemize}

\begin{figure}[ht]
    \centering
    \includegraphics[width=\columnwidth]{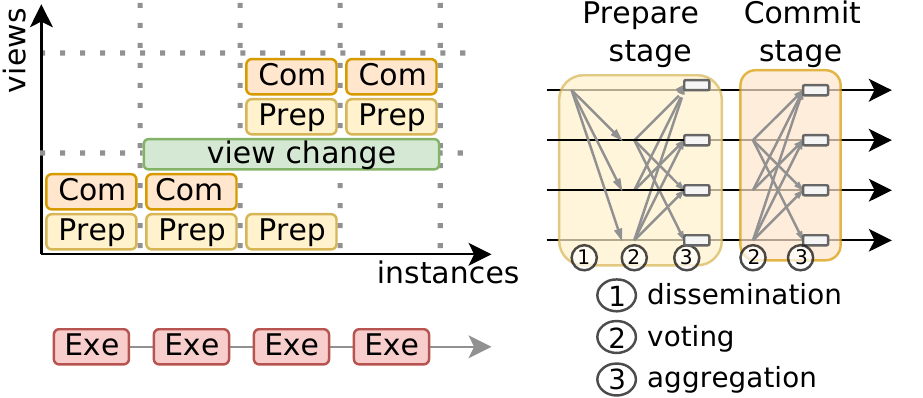}
    \caption{A simplified BFT SMR protocol model.}
    \label{fig:bft-simple}
\end{figure}

\subsection{BFT Simplified}


A BFT protocol implements \ac{SMR} in a Byzantine fault model.
To explain BFT protocols, we employ an abstract model (see \cref{fig:bft-simple}) that contains many common design aspects from BFT protocols, and we borrow some terminology from PBFT.
BFT protocols can be leaderless~\cite{borran2010leader, antoniadis2021leaderless} or work with multiple leaders~\cite{alqahtani2021bigbft, cong2018blockchain, stathakopoulou2019mir}, but most BFT protocol designs operate using a single leader. 

To implement \ac{SMR}, replicas must repeatedly agree on a block containing one or more inputs. 
We refer to one such agreement as an \textit{instance}. 
To reach agreement typically requires the successful execution of two or more \textit{protocol stages}.
The two protocol stages, {\smaller\textsc{prepare}} and {\smaller\textsc{commit}} of PBFT are shown in \cref{fig:bft-simple}.
Each protocol stage includes \textit{dissemination} of one or multiple proposals, \textit{voting} for a proposal by all replicas, and 
\textit{aggregation} of votes. 
In some protocols and stages, replicas only vote on whether aggregation was successful. 
Thus, dissemination may be omitted, as shown in the commit stage in \cref{fig:bft-simple}.
During aggregation, matching votes from different replicas are gathered to form a \textit{quorum certificate}: Quorum certificates contain votes from sufficiently many replicas to ensure that no two different values can receive a certificate; the value is now \textit{committed}.

After reaching an agreement in one instance, transactions in the decided (committed) block are ready for their \textit{execution} in all correct replicas. Subsequently, correct replicas reply back to the respective clients. 

In addition to instances, BFT protocols typically operate in logical views, where each \textit{view} describes one composition of replicas and may use a predefined leader or certain dissemination pattern.
While some protocols perform all stages of different instances in one view as long as agreement can be reached, as is shown in \cref{fig:bft-simple}, other protocols change the view for every stage or instance.
To change the view requires a \textit{view change} mechanism.
If not sufficiently many replicas reach agreement, \eg due to a faulty leader, the view change mechanism can synchronize replicas, replace the protocol leader, and eventually ensure progress by installing a leader under whom agreement instances finally succeed.
During view change, replicas exchange quorum certificates to ensure that agreements from previous views are continued in the next.


\section{Methodology}
\label{sect:methodology}

We create this survey by systematically reviewing a large body of relevant research work.
To find research work and determine its relevance, we conducted a systematic search, which we briefly explain in this section.
We performed a systematic search focusing on the scalability of BFT consensus.
Since we aim to increase transparency and reproducibility, we briefly present our search methodology to gather and select these research papers, which mainly originated from the field of distributed systems, and especially the BFT and blockchain communities.
 Our main ambitions are the following:
\begin{itemize}[leftmargin=1em, itemsep=.1em, parsep=.1em, topsep=.1em,partopsep=.1em]
 \item Find distinct approaches in the literature for improving the \emph{scalability of BFT consensus}.
 \item Identify and classify scalability-enhancing techniques to create a taxonomy for scalable BFT.
\end{itemize}

\subsection{Search Strategy} 
For literature research, we used Semantic Scholar~\cite{semantic_scholar}, whose API allows the automated download of a large number of references. 
We manually crosschecked with Google Scholar to ensure relevant papers were included.
We employed the following search queries without filters on the publication date: 

\begin{quote}
\underline{Search~1:} \\
\textit{Query} = Byzantine consensus scalability blockchain \\
\textit{Publication Type} = Journal or Conference \\
\textit{Sorted by Relevance (Citation Count)} = 250 most cited publications 
\end{quote}
 \begin{quote}
\underline{Search~2:} \\
\textit{Query} = Byzantine consensus scalability blockchain \\
\textit{Publication Type} = Journal or Conference \\
\textit{Sorted by Date} = 250 newest publications 
 \end{quote}
 
For each search query, on December 17, 2021, we downloaded all results (954 papers) and sorted them locally by relevance (Search~1) and by date (Search~2). 
The first search favors relevance (citation count) to ensure that we can cover the most influential works for the topic, while the second search favors publication date to ensure that we can also regard the most recent publications, which might not have been cited often enough due to their recency. 
With these searches, we cover all publications with at least 11 citations and further capture all publications that have been published since January 2021. 
By using two different sortings, we aim to include both the most relevant and most recent works. 
We found that the search engine could handle synonyms well, \eg papers would not evade our systematic search when using \enquote{agreement} instead of \enquote{consensus}.
After that, we combined the results of these queries and sanitized them for duplicates.
 
 

\subsection{Selection Criteria} 
Further, we employ a set of selection criteria that can be applied to determine if a paper found by the search queries is original research and \textit{potentially relevant}. This means it is selected for inclusion in the survey as long as no exclusion criteria apply. These criteria also encompass conditions used to assess the quality of a found paper. 
 
 \emph{Inclusion criteria} -- A paper is included if its contributions satisfy one of the following criteria:
 
 \begin{itemize}[leftmargin=1em, itemsep=.1em, parsep=.1em, topsep=.1em,partopsep=.1em]
\item The paper presents a novel technique for \textit{improving the scalability} of communication-based Byzantine consensus
\item The paper \textit{combines existing scalability-enhancing techniques}, or ideas, in a novel way, achieving better results than state-of-the-art protocols
 \end{itemize}

 \emph{Exclusion criteria} -- The exclusion process is applied \textit{after inclusion}.
 A paper is excluded if only one of the following criteria applies:
 
  \begin{itemize}[leftmargin=1em, itemsep=.1em, parsep=.1em, topsep=.1em,partopsep=.1em]
 	\item The paper is not a research paper (no practical reports, workshop invitations, posters, or other surveys).
 	\item The paper does not cover the scalability aspect sufficiently from a technical point of view or does not originate from the field of computer science (\ie not originating from the right `field', \eg we do not want papers from business informatics or social science).
 	\item The presented paper proposes a scalability mechanism but does not present a careful experimental evaluation showing how it can actually improve scalability (not enough validation).
 	\item The paper lacks relevance: while the paper presents some scalability mechanism(s), the presented mechanism is an incremental refinement of an earlier proposed mechanism.
 	\item The paper does not focus on communication-based Byzantine consensus but presents a `Proof-of-X' consensus variant.
 \end{itemize}

\subsection{Selection Procedure and Results} 
We selected papers as \textit{relevant} in the following way: In the first phase, we conducted a fast scan, in which a single assessor examined each paper only to check if the paper could be of potential interest. Moreover, in this step, only papers that obviously do not qualify are sorted out. For instance, if the field is not computer science, the paper's topic is not related to scalability, or the paper is not a research paper but a short abstract or workshop invitation.
In a second review round, all remaining papers are examined by two different reviewers to validate their relevance: A paper is \textit{relevant} iff at least one inclusion criterion applies, and none of the exclusion criteria applies. All papers are then tagged either \emph{relevant} or \emph{not relevant}.
In the final phase, assessor conflicts are resolved by involving a third assessor and discussing the disagreement.

In the end, $52$ papers have been selected as relevant for inclusion in our SoK paper.
Out of these, 13 papers are published in 11 different journals, 7 papers are published in online archives (\eg arXiv), and 32 are published in proceedings of 25 different conferences. 
The papers are spread over many conferences and journals, many not specific to the topic of blockchain. The most frequent publication channel, besides arXiv, is VLDB, with 3 papers.

\section{SoK: Scalability-enhancing Techniques}
\label{sect:scalability}

Improving the scalability of Byzantine consensus is an ongoing research field that
requires  the exploration, advancement, and combination of several methods and approaches.
In order to divide the different directions of approaches, we identified the following categories for scalability-enhancing techniques 
which aim to increase the scalability (and efficiency) of Byzantine consensus:

\begin{itemize}[leftmargin=1em, itemsep=.1em, parsep=.1em, topsep=.1em,partopsep=.1em]
\item \textit{Communication topologies and strategies}. How can the communication flow be optimized? Increasing the communication efficiency might require a suitable \textit{communication topology}, \eg tree-based or overlay / gossip-based. 
\item \textit{Pipelining}. What benefit can be achieved by employing pipelining techniques for parallel executions of agreement instances?
\item \textit{Cryptographic primitives}. How can \textit{suitable cryptographic primitives} serve as building blocks for new scalability-enhancing techniques?
\item \textit{Independend groups}. How can transactions be ordered and committed \textit{in independent groups}, \eg through sharding?
\item \textit{Consensus committee selection}. How are roles in achieving agreement distributed among network nodes? Does one (or multiple) flexibly selected \textit{representative committee(s)} decide?
\item \textit{Hardware support}. How can we improve consensus efficiency using \textit{\acp{TEE}}?
\end{itemize}

\begin{figure}
    \centering
    \begin{tikzpicture} 
    \begin{axis}[ 
    font= \footnotesize,
     ylabel={papers found}, 
     xticklabels from table={data/paper-per-category.txt}{category},   
        x tick label style={rotate=30,anchor=east,  xshift=10pt, yshift=-4pt,   font= \footnotesize},
     ybar=2pt,  
     bar width=6pt,
    height=3.67cm,
       xtick=data, 
           nodes near coords={
        \pgfmathprintnumber[precision=0]{\pgfplotspointmeta}
       },
       ytick = {5, 10, 15, 20, 25},
        ymin=0,
        ymax=30,
        xmin=0.5,
        xmax=8.5,
    ymajorgrids=true,
    yminorgrids=true,
    minor grid style={dashed,gray!10},
    minor tick num=1,
    legend style={at={(1, 1.08)},
    legend columns = 5,
    legend cell align=left
    }
    ] 
      \addplot 
      [draw = blue,
        fill = blue!30!white]   
        table[ 
          x=position, 
          y=papers   
          ] 
      {data/paper-per-category.txt}; 
    \end{axis}
\end{tikzpicture} 
\vskip -0.1 cm
    \caption{For each category: How many papers used at least one scalability-enhancing technique fitting to this category.}
    \label{fig:papersPerCategory}
\end{figure}
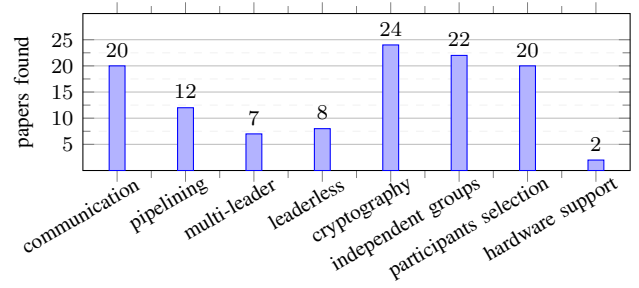

In~\cref{fig:papersPerCategory}, we show an overview of how many papers we found to fit into each category.

\subsection{Communication Topologies and Strategies} 
\label{sec:communication-topologies}

Improving the scalability of coordinating agreement between a large number of $n$ nodes requires avoiding bottleneck situations, such as burdening too much communication effort on a single leader.
An asymmetric utilization of network or computing resources hinders scalability, as observed in many traditional BFT protocols resembling PBFT. 
Instead, one of the main goals of the \enquote{blockchain generation} of BFT protocols is to distribute the communication load among replicas as evenly as possible. 
For instance, some approaches try to balance network capabilities under a single leader by using message forwarding and aggregation along a communication tree~\cite{kogias2016enhancing, neiheiser2021kauri} or gossip~\cite{buchman2018latest}, while others rely on utilizing multiple leaders~\cite{stathakopoulou2019mir, alqahtani2021bigbft} or even work fully decentralized without a leader~\cite{crain2018dbft, voron2019dispel, crain2021red, antoniadis2021leaderless}. 

Another reason why the well-known PBFT protocol does not scale well is its all-to-all broadcast phases which make the network topology a \textit{clique}, which is usually always a bad choice for scalability as it means an incurred $O(n^2)$ number of messages (and necessary message authenticators). Unsurprisingly, this design is not an ideal fit for a large network size $n$ and was later replaced by protocols that only require linear message complexity by collecting votes~\cite{gueta2019sbft, hotstuff19}. 

In the following, we review and explain different approaches for improving communication flow in BFT protocols categorized by the network topology they employ.

\subsubsection{Star-based} 

\begin{figure}[t]
  \centering
\begin{subfigure}[h]{\columnwidth}
  \centering
    \includegraphics[width=0.35\textwidth]{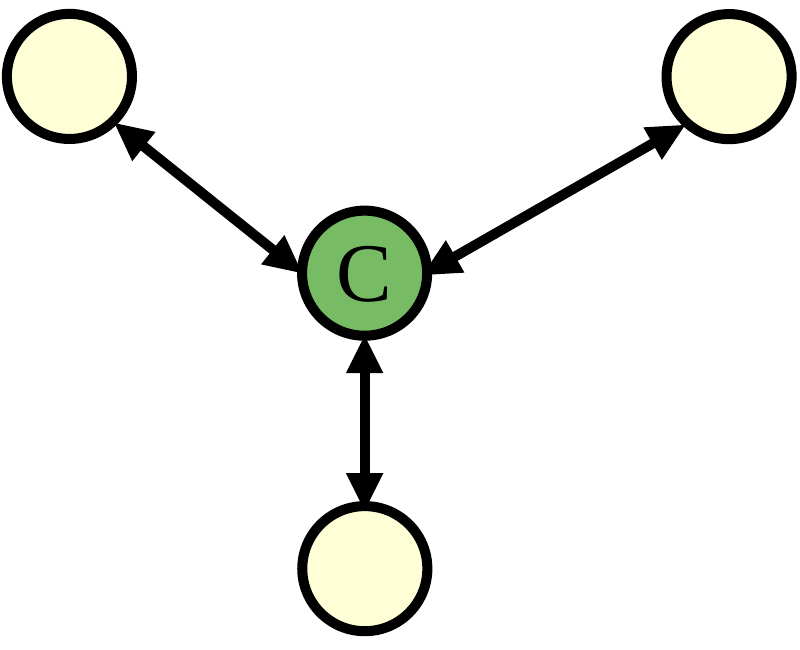}
    \caption{Star topology: A collector (C) collects votes and distributes quorum certificates (in Figure~\ref{fig:scalability:hotstuff-protocol}, replica $r_0$ has this role).}
    \label{fig:scalability:star-topology}
\end{subfigure}
\hspace{0.4cm}
\begin{subfigure}[h]{\columnwidth}
  \centering
    \includegraphics[width=\textwidth]{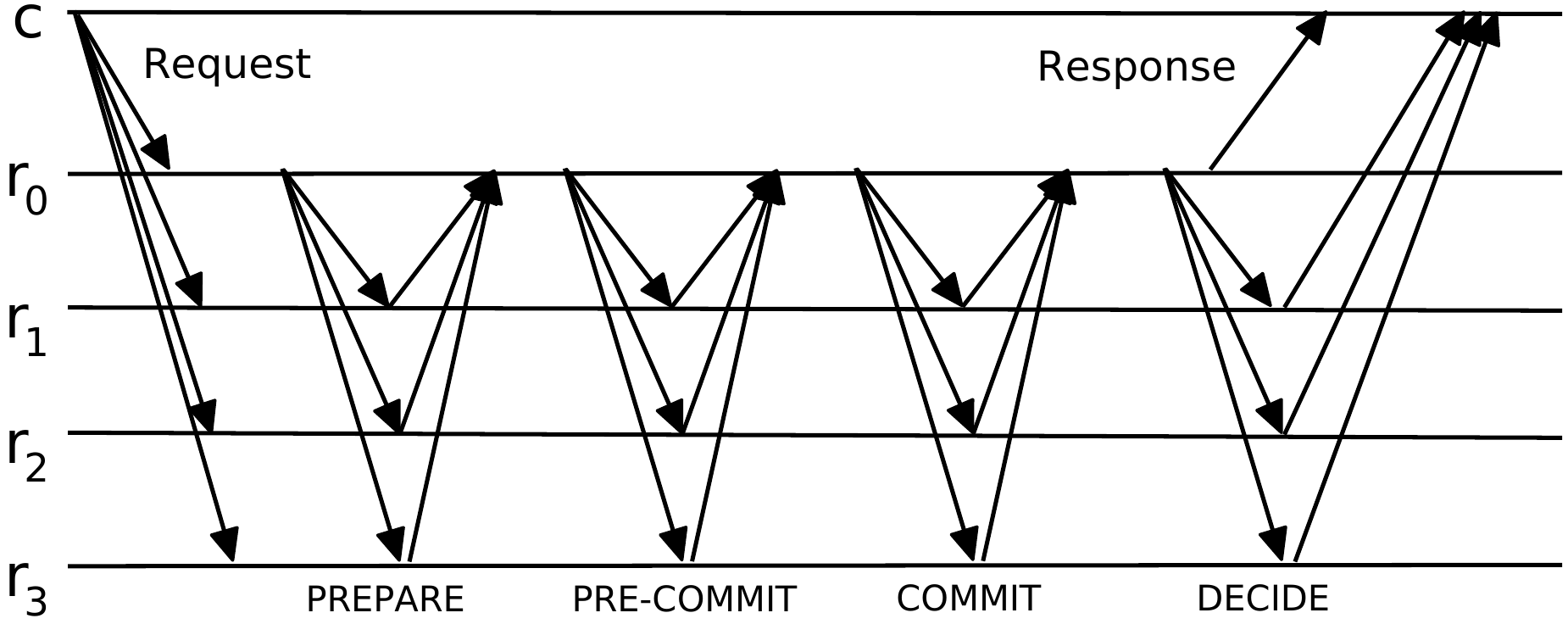}
    \caption{HotStuff has linear communication complexity: The leader collects the votes from the other replicas and distributed the quorum certificates. Aggregation can be achieved through threshold signatures.}
    \label{fig:scalability:hotstuff-protocol}
\end{subfigure}
\caption[]{Linear communication over a star-based network topology on the example of HotStuff.}
\label{fig:scalability:communication:star} 
\end{figure}

Recently, many BFT protocols have proposed to employ linear communication complexity by employing a star-based communication topology~\cite{hotstuff19, gueta2019sbft, jalalzai2018window, jalalzai2019proteus, gupta2021poe, gelashvili2021jolteon, dang2019towards}. We briefly explain this idea and its limitations on the example of HotStuff~\cite{hotstuff19} (which is illustrated in~\cref{fig:scalability:communication:star}).

The HotStuff leader acts as a \textit{collector}, which gathers votes from other replicas, and, as soon as a quorum of votes is received, creates a \textit{quorum certificate} that can convince any node to progress to the next protocol stage. The rather costly \textit{view-change} subprotocol of PBFT can be avoided by adding a further protocol stage.
Thus, agreement in HotStuff requires four protocol stages ({\smaller\textsc{prepare}, \textsc{pre-commit}, \textsc{commit},} and {\smaller\textsc{decide}}) shown in \cref{fig:scalability:hotstuff-protocol}.

Important for scalability is that the cost of transmitting message authenticators can be reduced by a simple aggregation technique: The leader compresses $n-f$ signatures into a single threshold signature of fixed size. Because the threshold signature scheme uses the quorum size as a threshold, a valid threshold signature implies a quorum of replicas is among the signers. The threshold signature has the size of $O(1)$ -- a significant improvement over letting the leader transmit $O(n)$ individual signatures.
HotStuff also employs pipelining and leader rotation which are addressed in \cref{sect:pipelining}.

As we can see in~\cref{fig:scalability:hotstuff-protocol}, the communication flow still is imbalanced:
Every follower replica communicates only with the leader, but the leader still has to communicate with all other replicas. 
This bottleneck can be alleviated by tree-based or randomized topologies, as we explain next.


\begin{figure}[t]
  \centering
\begin{subfigure}[h]{\columnwidth}
  \centering
    \includegraphics[width=0.35\textwidth]{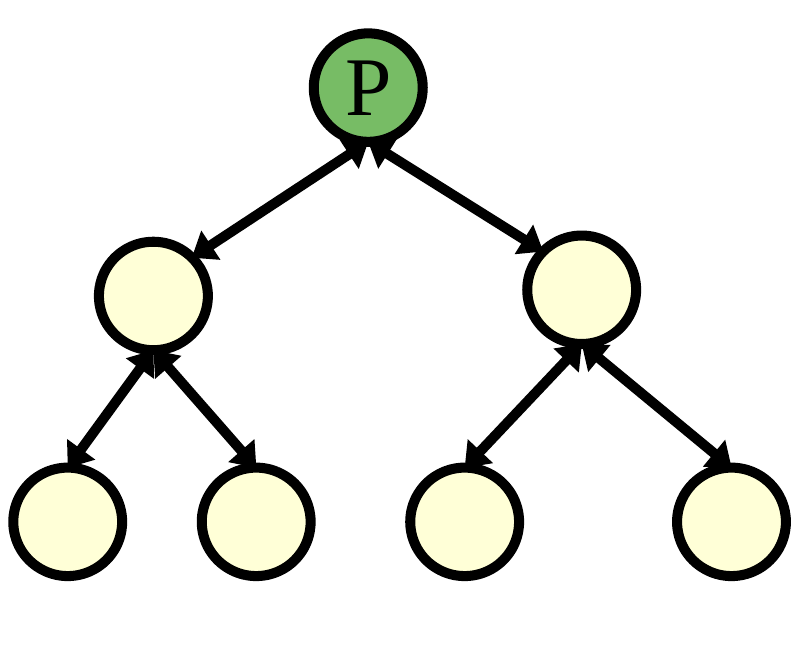}
    \caption{Tree topology: The proposer (P) disseminates messages  and collects votes along a tree. 
    }
    \label{fig:scalability:tree-topology}
\end{subfigure}
\begin{subfigure}[h]{\columnwidth}
  \centering
    \includegraphics[width=\textwidth]{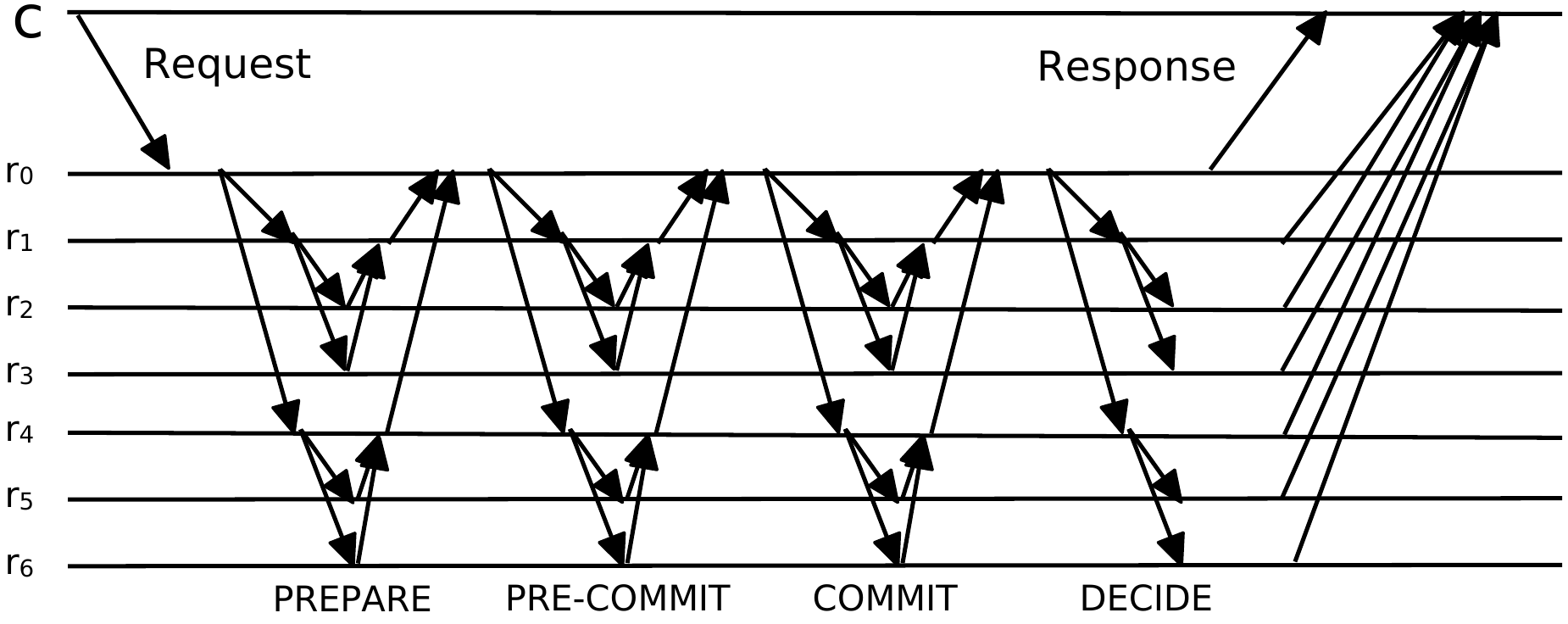}
    \caption{In Kauri, the leader disseminates a proposal along a communication tree. The inner nodes forward messages to their children and collect and aggregate their children's votes, to be returned to the parent.}
    \label{fig:scalability:kauri-protocol}
\end{subfigure}
\caption[]{Communication over a tree-based network topology on the example of Kauri.}
\label{fig:scalability:communication:tree} 
\end{figure}

\subsubsection{Tree-based} A few novel BFT protocols employ a tree-based communication topology~\cite{kogias2016enhancing, li2021scalable, liu2018scalable, neiheiser2021kauri}. This has the advantage that the leader is relieved of the burden of being the sole aggregation and dissemination point for votes and the generated quorum certificates.
The tree-based communication pattern can be introduced into either the PBFT algorithm, as done in ByzCoin~\cite{kogias2016enhancing}, or HotStuff, as done in Kauri~\cite{neiheiser2021kauri}, without requiring significant changes to the protocol.
ByzCoin was the first to show the potential of this topology, while Kauri later showed how to overcome the shortcomings of this approach.

The tree-based communication topology raises several problems, such as added latency, compared to the star topology and the necessity to react to failures of internal and leaf nodes.
In the following, we briefly explain these challenges and the solutions proposed.
\cref{fig:scalability:communication:tree} shows the communication pattern in Kauri.
Compared to the pattern in~\cref*{fig:scalability:communication:star}, it is clear that each level in the tree increases latency. 
To address this issue, Kauri only considers trees of height 3, as shown in the Figure.
Additionally, in the star topology, only leader failure is critical and typically requires a view change. 
In the tree topology, the failure of an internal node prevents the aggregation of votes from its children and thus requires reconfiguration.
Further, the failure of a leaf node may cause its parent to wait indefinitely for its contribution.
Kauri introduces a timeout for aggregation to handle leaf failure. 
To address the failure of internal nodes, Kauri proposes a reconfiguration scheme, which guarantees to find a correct set of internal nodes, 
given that failures lie below a threshold $k$.
Otherwise, Kauri falls back on a star-based topology.
The threshold $k$ here depends on the tree layout but lies at $\sqrt{n}$ for a balanced tree.

To achieve competitive throughput and utilize the available bandwidth despite the latency added through the tree overlay and through timeouts on leaf nodes, Kauri relies on pipelining. 
We will see more details on how pipelining can serve as a means for increasing scalability in~\cref{sect:pipelining}.

\begin{figure}[t]
  \centering
\begin{subfigure}[h]{\columnwidth}
  \centering
  \centering
    \includegraphics[width=0.35\textwidth]{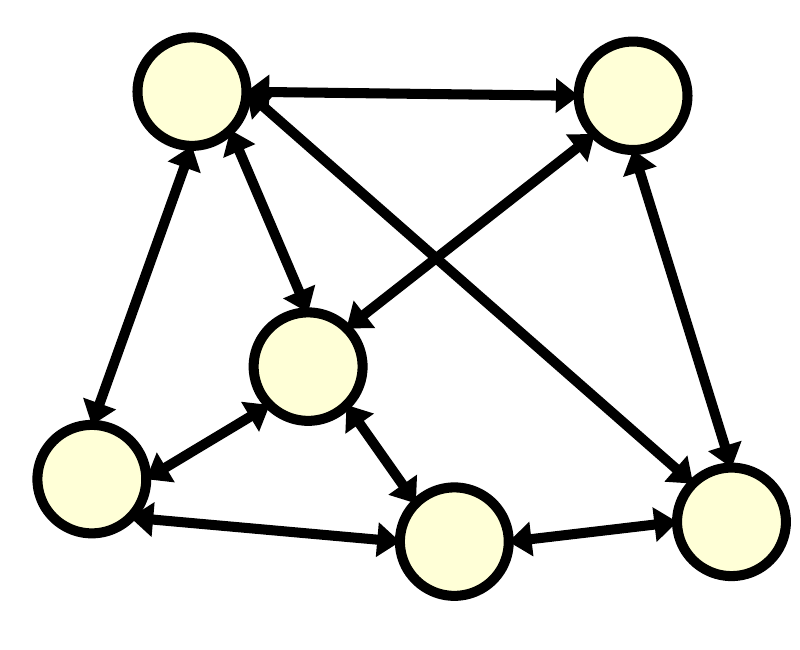}
    \caption{Randomized topology: Gossip between randomly chosen, connected nodes that are called \textit{neighbors}. 
    }
    \label{fig:scalability:gossip}
\end{subfigure}
\hspace{0.4cm}
\begin{subfigure}[h]{\columnwidth}
  \centering
    \includegraphics[width=\textwidth]{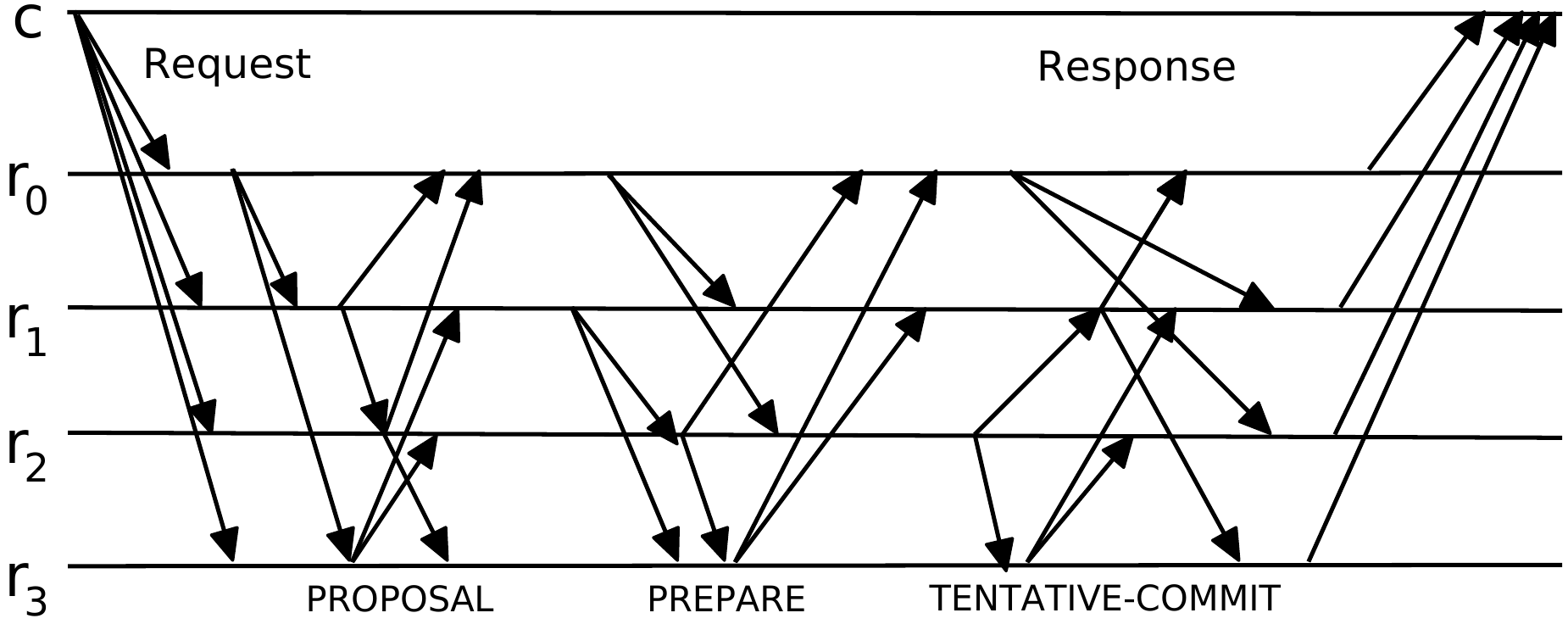}
    \caption{In Gosig, replicas communicate with a \textit{fanout} (\ie number of randomly selected neighbors) to disseminate a block or votes for a block within each protocol stage. Each instance is started by a randomly selected leader.}
    \label{fig:scalability:gosig-protocol}
\end{subfigure}
\caption[]{A randomized communication strategy through gossip on the example of Gosig~\cite{li2020gosig}.}
\label{fig:scalability:communication:gossip} 
\end{figure}

\subsubsection{Gossip and Randomized Topology} Furthermore, many recent BFT protocols rely on gossip, or randomized communication topologies, for instance 
Internet Computer Consensus~\cite{camenisch2022internet}, Tendermint~\cite{cason2021design}, Algorand~\cite{gilad2017algorand}, Gosig~\cite{li2020gosig}, scalable and leaderless Byz consensus~\cite{lim2014scalable}, Avalanche~\cite{rocket2020avalanche} and  RapidChain~\cite{zamani2018rapidchain}.

As shown earlier, deterministic leader-based BFT consensus protocols are much affected by the high communication costs of broadcasts, \ie if the leader disseminates a block of ordered transactions to all other nodes. An idea that enhances scalability is to relieve the leader and shift this burden equally to all nodes by disseminating messages through gossiping over a randomized overlay network – which was recently proposed by protocols like Algorand~\cite{gilad2017algorand}, Tendermint~\cite{cason2021design}, or Gosig~\cite{li2020gosig}. 

When using gossip, the leader proposes a message $m$ to only a constant number of $k$ other, randomly chosen nodes. Nodes that receive $m$ forward the message to their own randomly chosen set of connected neighbors, just like spreading a rumor with high reliability in the network. Note that this technique is probabilistic, and the probability of success (as well as the propagation speed) depends on the fanout parameter $k$, the number of hops, and the number of Byzantine nodes present in the system.
Gossip allows for the  communication burden to be lifted from the leader, leading to a fairer distribution of bandwidth utilization, where each node communicates only with its $O(k)$ neighbors, thus improving scalability. 

\cref{fig:scalability:communication:gossip} shows the communication pattern of Gosig~\cite{li2020gosig}.
Similar to the tree-based overlay, the communication pattern leads to increased latency.
As Kauri, Gosig uses pipelining to mitigate this latency.
Different from Kauri, randomized topologies do not require reconfiguration in case of failures.


Random overlay network topologies can also be utilized in leaderless BFT protocols. Avalanche~\cite{rocket2020avalanche} is a novel, leaderless BFT consensus protocol that achieves consensus through a metastability mechanism that is inspired by gossip algorithms. In Avalanche, nodes build up confidence in a consensus value by iterative, randomized mutual sampling. Each node queries $k$ randomly chosen other nodes in a loop while adopting the value replied by an adjusted majority of nodes so that eventually, correct nodes are being steered towards the same consensus value. In this probabilistic solution, nodes can quickly converge to an irreversible state, even for large networks.


\subsection{Pipelining}
\label{sect:pipelining}



Pipelining is a technique that allows replicas to run multiple instances of consensus concurrently, and it is employed with a special focus on improving the scalability of the BFT protocol, \eg in HotStuff~\cite{hotstuff19}, Kauri~\cite{neiheiser2021kauri}, BigBFT~\cite{alqahtani2021bigbft}, SBFT~\cite{gueta2019sbft}, \ac{PoE}~\cite{gupta2021poe}, ResilientDB~\cite{gupta2020resilientdb}, 
Dispel~\cite{voron2019dispel}, RapidChain~\cite{zamani2018rapidchain}, Hermes~\cite{jalalzai2021hermes}, RCC~\cite{rcc2021gupta}, Gosig~\cite{li2020gosig}, Narwhal-HotStuff/Tusk~\cite{danezis2022narwhal}, and Jolteon/Ditto~\cite{gelashvili2021jolteon}.

Further, pipelining  boosts performance and scalability because replicas  maximize their available resource utilization. For instance, the available bandwidth can be better utilized if data belonging to future consensus instances can be disseminated while  replicas still wait to collect votes from previous, ongoing consensus instances.
This means pipelining can systematically decrease the time replicas spend in an idle state, \eg waiting to collect messages from others. 
Gosig~\cite{li2020gosig}, for example, pipelines both the BFT protocol, as committed nodes can start a round while still forwarding gossip messages on the previous one, and the gossip layer, by decoupling block and signature propagation.
With pipelining, a replica can participate in multiple consensus stages simultaneously, 
and thus help to overcome \textit{limitations} for system performance, such as \textit{reusing quorum certificates} for multiple consensus stages in HotStuff or \textit{mitigating tree latency} in Kauri.

\subsubsection{Out-of-Order Processing}
A basic variant of pipelining was also introduced with PBFT: The leader can build a pipeline of concurrent consensus  instances by starting multiple instances for a specific allowed range defined through a low and high watermark. 
Within this bound, pipelining is \textit{dynamic}, allowing the protocol to start fewer or more concurrent instances, based on the load.
PBFT employs \textit{out-of-order} processing (see~\cref{fig:pipelinePBFT}), which means that replicas in the same view vote on and commit in all allowed consensus instances without waiting for preceding instances to complete first. Yet, in this case, executions must be delayed until all transactions within lower-numbered consensus instances have been executed.

Similar to PBFT,  the SBFT protocol~\cite{gueta2019sbft} allows the parallelization of blocks, in which multiple instances of the protocol can run concurrently,
but it additionally introduces an adaptive learning heuristic to \textit{dynamically optimize its block size}. 
This heuristic employs a configurable maximal recommended parallelism parameter, the number of currently ongoing, pending blocks, and the number of outstanding client requests. SBFT's heuristic strives to distribute and balance pending client transactions evenly among the recommended number of parallel executing consensus instances, and new requests are not required to wait when currently no decision block is being processed. 

Moreover, \ac{PoE}~\cite{gupta2021poe} is a new BFT design employed within ResilientDB~\cite{gupta2020resilientdb} for achieving high throughput using a multi-threaded pipelined architecture. 
Its core innovation is to combine out-of-order processing and dynamic pipelining, as introduced in PBFT, with \textit{speculative executions}, where transactions are executed before agreement.
Since \ac{PoE} allows replicas to execute requests speculatively, it introduces a novel view-change protocol that can rollback requests.

\subsubsection{Chain-based Pipelining \& QC Reusage}

In chain-based protocols like HotStuff, a block is committed after having passed through several protocol stages, as shown in \cref{fig:scalability:hotstuff-protocol}. In each stage, the leader collects votes from the other replicas, aggregates these in a quorum certificate (QC), and disseminates this QC to all as proof of completing the respective protocol stage. 
HotStuff incorporates the idea of pipelining by making each protocol stage into a pipelining stage (see \cref{fig:pipelineHS}), and thus allowing a QC to certify advanced stages of previous, concurrently executing consensus instances, \eg a {\small\textsc{decide}} QC of instance $i$ can act as a {\small\textsc{commit}} QC of instance $i+1$ and {\small\textsc{pre-commit}} QC of instance $i+2$. 
Pipelining in HotStuff is \textit{logical}, meaning that concurrent instances are using the same messages.
The number of pipelining stages in HotStuff equals its number of protocol stages, namely four. 
Narwhal-HotStuff and Tusk~\cite{danezis2022narwhal} as well as Jolteon and Ditto~\cite{gelashvili2021jolteon} built on HotStuff and use the same pipelining mechanism.

\subsubsection{Multiplexing Consensus Instances per Protocol Stage}
\begin{figure}[t]
  \centering
  \begin{subfigure}[h]{\columnwidth}
  \centering
    \includegraphics[width=\textwidth]{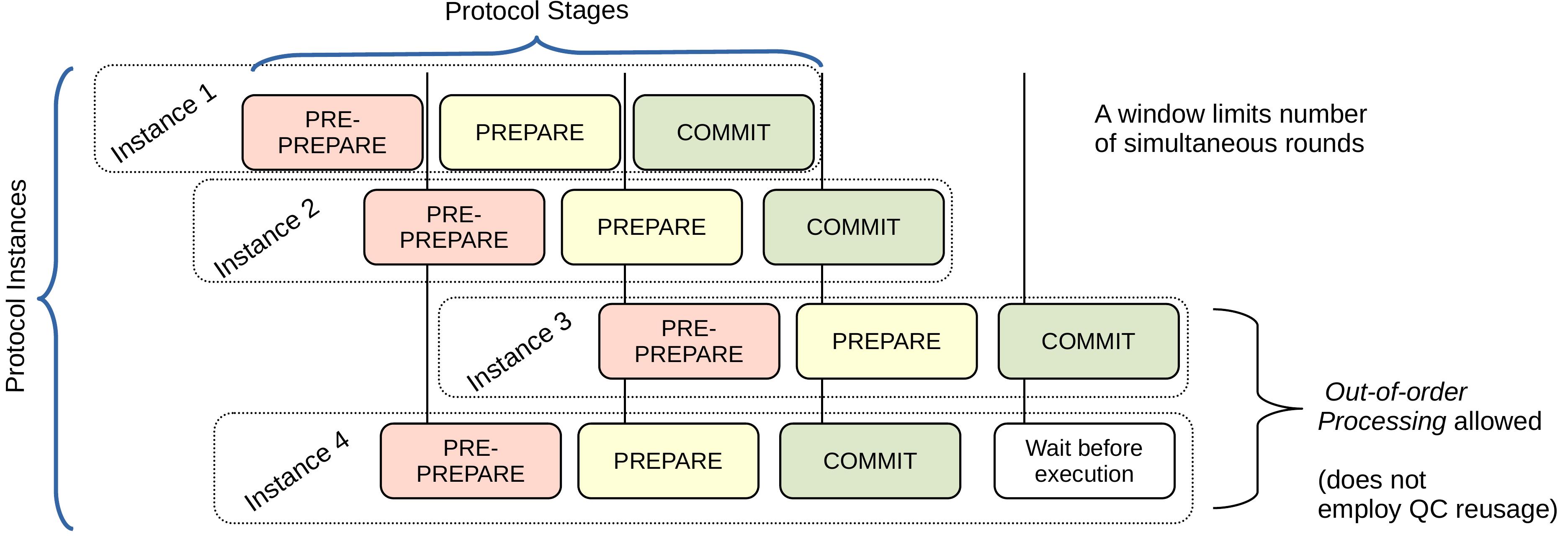}
    \caption{\textit{Out-of-Order Processing} (here on the example of PBFT): The leader can start multiple instances for a given window of allowed consensus instances concurrently. QCs are not being reused. 
    }
    \label{fig:pipelinePBFT}
\end{subfigure}
\begin{subfigure}[h]{\columnwidth}
  \centering
    \includegraphics[width=\textwidth]{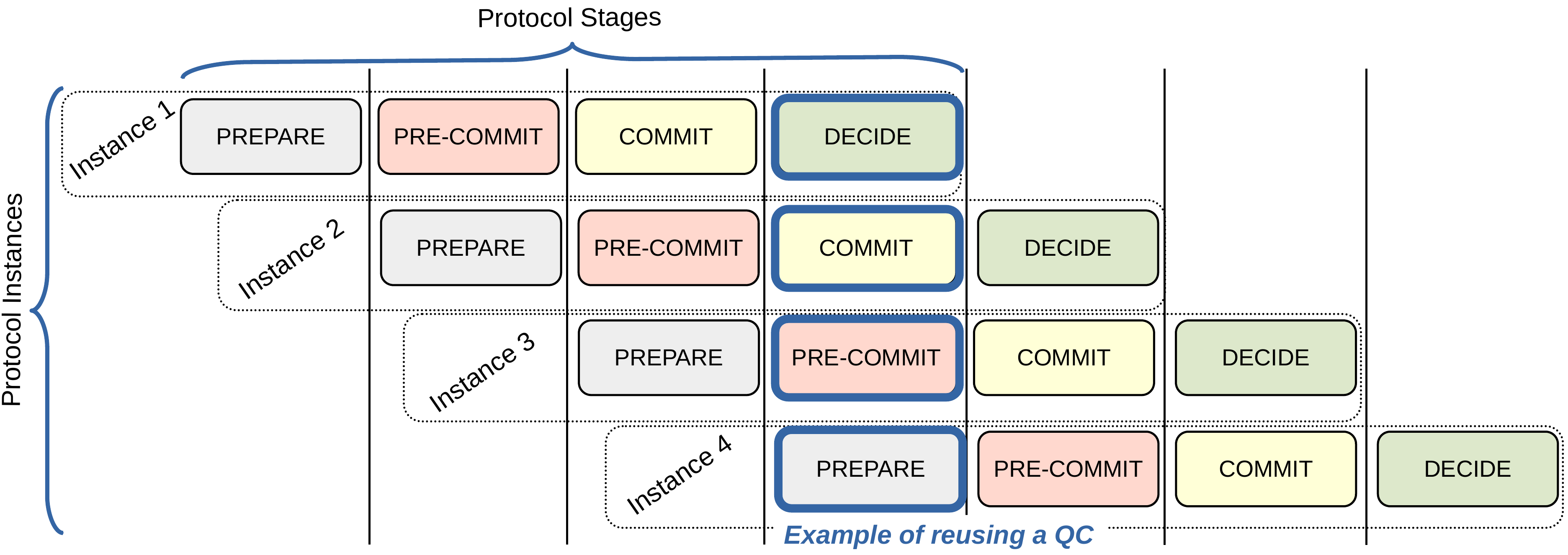}
    \caption{\textit{Chain-based Pipelining} uses one pipeline stage per protocol stage (this example shows HotStuff). QCs can be reused to verify incremental protocol stages of concurrently running instances. 
    }
    \label{fig:pipelineHS}
\end{subfigure}
\begin{subfigure}[h]{\columnwidth}
  \centering
    \includegraphics[width=\textwidth]{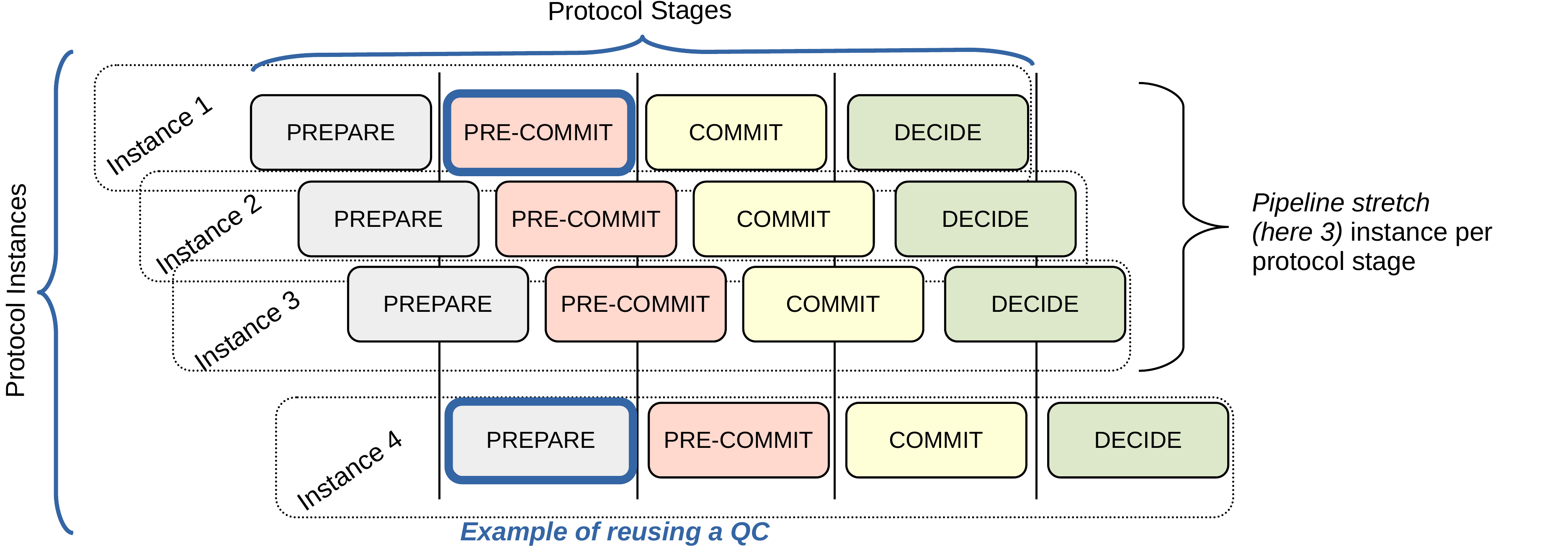}
    \caption{\textit{Multiplexing Consensus Instances per Protocol Stage}: A \textit{pipelining stretch} number of consensus  instances can be concurrently started per protocol stage (as done in Kauri). QCs can be reused.}
    \label{fig:pipelineKauri}
\end{subfigure}
\caption[]{Pipelining variations of single leader BFT SMR.}
\label{fig:scalability:communication:pipelining} 
\end{figure}

Chain-based pipelining can be further optimized by multiplexing consensus instances per protocol stage, as done by Kauri. For this purpose, 
Kauri refines the communication flow of HotStuff by proposing a tree topology with a fixed-sized fanout (number of communication partners). This load-balancing technique relieves the necessary uplink bandwidth (and thus also the time needed to send data) for message distribution at the root (leader). 
But at the same time, the communication tree increases the number of communication steps necessary for reaching agreement and thus impacts the overall latency of the SMR protocol. In Kauri, a more sophisticated pipelining method than in HotStuff acts as a solution to sustain performance despite this higher number of communication steps: The number of concurrently run consensus instances is decoupled from the fixed number of protocol stages by introducing a \textit{stretching factor}, which is a parameterizable multiplicative to the fixed pipelining depth of HotStuff. For instance, if a pipeline stretch of $3$ is used, the Kauri leader can simultaneously start 3 consensus instances every time a HotStuff leader would start a single instance, leading to a total of $12$ concurrent consensus instances instead of 4 (like in HotStuff) once the pipeline is filled. In contrast to PBFT, the pipeline stretch in Kauri is static and cannot dynamically adapt to the load, and instances cannot complete out-of-context (see \cref{fig:pipelineKauri}).

\subsubsection{Pipelining in Multi-leader and Leaderless Protocols} Many BFT protocols rely on the idea of having not only one but several leaders that can concurrently start agreement instances (\eg~\cite{alqahtani2021bigbft, cong2018blockchain, stathakopoulou2019mir}). This design benefits scalability because it brings a fairer distribution of the task of broadcasting block proposals and can prove a suitable solution for the leader bottleneck problem. Yet, it also introduces novel challenges, in particular when it comes to the \textit{coordination of leaders}, to prevent conflicts (\ie identical transaction being proposed by multiple leaders in different blocks) and also to maintain \textit{liveness} for all transactions in the presence of Byzantine leaders.  
To serve as an example, Mir-BFT~\cite{stathakopoulou2019mir} is a multi-leader protocol in which several leaders propose blocks independently and in parallel, by employing a mechanism that rotates the assignment of a partitioned transaction hash space to the leaders. Another example is 
RapidChain~\cite{zamani2018rapidchain}, which is a BFT protocol that uses sharding and allows a shard leader to propose a new block while re-proposing the headers of the yet uncommitted blocks, thus creating a pipeline to improve performance. 

BigBFT~\cite{alqahtani2021bigbft} is a pipelined multi-leader BFT design that aims to overcome the
leader bottleneck of traditional BFT protocols by partitioning the space of sequence numbers (consensus instances) among selected leaders (coordination stage) and thus
allowing multi-leader executions. Further, it logically separates the coordination stage from the dissemination and voting of new blocks. Pipelining is employed in two ways.
Different leaders perform their instances concurrently.
Additionally, the next coordination stage is run concurrently with the last agreement stages.
Similarly, RCC~\cite{rcc2021gupta} concurrently executes multiple instances with different leaders.
Instead of a coordination stage, RCC uses a separate, independent instance of a BFT protocol for coordination.

A key innovation of Dispel~\cite{voron2019dispel} is its \textit{distributed pipeline} which builds on previous work on leaderless BFT consensus~\cite{crain2018dbft}. 
In contrast to centralized pipelining (where the decision, when to create a new agreement instance, is only up to the leader(s)), 
all Dispel replicas can locally decide whether to start new consensus instances based on their available system resources like network bandwidth or memory. Dispel builds a consensus pipeline by creating pipelining stages for \textit{individual tasks} within running consensus that consumes different system resources. In particular, it employs four stages: network reception, network transmission, CPU-intensive hash, and latency-bound consensus~\cite{voron2019dispel}. The idea of Dispel is that a replica can maximize its own resource utilization by executing four consensus instances, one for each stage, concurrently.
 We cover
 \textit{leaderless  protocols} in more detail later in \cref{randomized-sampling}.

\subsection{Cryptographic Primitives}
\label{sec:crypto}

Efficient cryptography schemes can increase a protocol's scalability.
We identified several approaches, most aimed at reducing the communication complexity via message aggregation: multi-signatures (\cref{sec:crypto:aggregation}), 
threshold signatures (\cref{sec:crypto:threshold}), 
secret sharing (\cref{sec:crypto:secretsharing}), 
erasure coding (\cref{sec:crypto:erasurecoding}), 
as well as efficiently selecting a committee of nodes using \acp{VRF} (\cref{sec:crypto:vrf}). 

%

\subsubsection{Multi-Signatures}
\label{sec:crypto:aggregation}
Seven papers that match our criteria aggregate messages using multi-signatures: Kauri~\cite{neiheiser2021kauri}, Gosig~\cite{li2020gosig}, ByzCoin~\cite{kogias2016enhancing}, Musch~\cite{jalalzai2018window}, AHL~\cite{dang2019towards}, BigBFT~\cite{alqahtani2021bigbft}, and RepChain~\cite{huang2021repchain}.
A multi-signature allows $n$ participants, who want to jointly sign a common message $m$, to create a signature $\sigma$ of $m$ so that verification can confirm that all $n$ participants have indeed signed $m$.
This can be achieved by aggregating multiple signatures, \eg via multiplication, resulting in a multi-signature whose combined size and verification cost is comparable to that of an individual signature~\cite{boneh2018multisigbls}.

In BFT systems, multi-signatures are often used to combine the votes of multiple participants to reduce the message complexity and the memory overhead of the protocol from quadratic to linear~\cite{li2020gosig, neiheiser2021kauri, alqahtani2021bigbft, jalalzai2018window}.
A replica casts its vote by adding its partial signature share to the multi-signature, and the voting concludes once a quorum of signature shares has been received~\cite{neiheiser2021kauri}.
Most protocols make use of asynchronous non-interactive multi-signatures, \ie BLS multi-signatures~\cite{boneh01bls, boneh2018multisigbls}.
Replicas aggregate their signature share in one small multi-signature.
This reduces the signature size compared to individual replica signatures; however, the computation takes longer compared to, \eg ECDSA signatures.
Multi-signatures thus offer accountability, as it can be checked who signed a message.
Further, they have the advantage that the order of signatures can be arbitrary, making them resilient against adaptive chosen-player attacks~\cite{li2020gosig}.
A node can sign the message multiple times, \eg when it receives the same message from different nodes during gossip communication.
To ensure correct verification, the number of times each node signs the message is tracked~\cite{li2020gosig}.

Signatures are collected during the voting phase, \eg during gossip as in Gosig or during vote aggregation phases as in Kauri, where each replica receives votes from its children in the communication tree, and once a quorum of signatures is collected, the replica enters the next phase. 
BigBFT~\cite{alqahtani2021bigbft} is a multi-leader protocol where each leader proposes client requests in a block in instance $r$. 
All blocks are signed by all other nodes during the vote phase upon reception.
The set of $n-f$ votes for a block is then aggregated in a multi-signature, which is piggybacked onto the proposed block in the next instance $r+1$ to commit the block of instance $r$.
BigBFT avoids the communication bottleneck of single-leader protocols and reaches consensus in two communication rounds by pipelining blocks using message aggregation and multi-signatures to reduce message complexity. 
In Musch~\cite{jalalzai2018window}, 
nodes exchange and aggregate signed hashes of blocks to create one collective signature instead of repeating multiple replica signatures, thus keeping the signature size constant.
If not enough signature shares can be collected, a view change is performed, which also employs aggregated multi-signatures.

Another scheme that is used to aggregate signatures is the collective signing protocol CoSi~\cite{syta16cosi}, which can effectively aggregate a large number of signatures.
It is based on Schnorr multi–signatures and, contrary to the non-interactive BLS multi-signatures, it requires a four-phase protocol run over two round-trips to generate a CoSi multi-signature.
The participants can be organized in communication trees for efficiency and scalability, as discussed in \cref{sec:communication-topologies}.
A node can request a statement, \ie a request, to be signed by a group of witnesses, \ie the replicas, and the collective signature attests that the node, as well as the witnesses, have observed this request.
It is used in ByzCoin~\cite{kogias2016enhancing} to reduce the cost of the underlying PBFT's prepare and commit phases, as well as in RepChain~\cite{huang2021repchain}, to enable efficient cross-shard transactions without requiring multiple individual signatures.
However, the security of two-round multi-signatures has been shown to be compromised~\cite{drijvers2019multisigsecurity}.

\subsubsection{Threshold Signatures}
\label{sec:crypto:threshold}

A total of nine papers utilize threshold signatures in their protocol design: Jolteon/Ditto~\cite{gelashvili2021jolteon}, Saguaro~\cite{amiri2021saguaro}, ICC~\cite{camenisch2022internet}, Narwhal/Tusk~\cite{danezis2022narwhal}, \ac{PoE}~\cite{gupta2021poe}, HotStuff~\cite{hotstuff19}, SBFT~\cite{gueta2019sbft}, Cumulus~\cite{gai2021cumulus}, and Dumbo~\cite{guo2020dumbo}.
In threshold cryptosystems, participants each have a public key and a share of the corresponding private key: in $(t, n)$-threshold systems, at least $t$ partial shares from participants are needed to decrypt or sign a message.
Participants can sign a message with their secret key share, generating a signature share.
This signature share can be verified or combined with others into an aggregated signature, which can again be verified.
This makes threshold signatures a special case of multi-signatures, where instead of all participants ($n$-out-of-$n$) only a subset ($t$-out-of-$n$, $t < n$) have to participate.
BLS multi-signatures can be transformed into threshold signatures; however, the implementation of threshold signature schemes is more complex, and multi-signatures require less computation~\cite{gueta2019sbft}.

Threshold signatures are, therefore, often used in BFT systems for aggregation of messages, similar to multi-signatures: a block or vote message is signed by a quorum of nodes (commonly with a threshold $t = 2f+1$), the quorum's shares are combined in one authenticator, and the signature share is used as a replica's vote to confirm consensus~\cite{amiri2021saguaro, camenisch2022internet, gelashvili2021jolteon, gupta2021poe, hotstuff19}, or for example to create a quorum certificate for side-chain checkpoints as in Cumulus~\cite{gai2021cumulus}.
BLS signatures can be used for this as well, as done by ICC~\cite{camenisch2022internet} and \ac{PoE}~\cite{gupta2021poe}: either by using the standard BLS scheme with a secret key shared amongst all participants, which creates unique and compact signatures, or by using BLS multi-signatures, where a signature share is a BLS signature which then gets combined into a new signature on an aggregate of the individual public key.
\ac{PoE} can use threshold signatures or message authentication codes depending on the number of participants in the network.

Threshold signatures can be used just as multi-signatures to split one phase of high-complexity broadcast communication into two phases of linear communication complexity.
Dumbo~\cite{guo2020dumbo} introduces two new asynchronous atomic broadcasting protocols.
The first protocol, Dumbo1, reaches asymptotical efficiency, improving upon the design of threshold encryption and asynchronous common subsets of HoneyBadgerBFT~\cite{miller2016honey}.
The second protocol, Dumbo2, further reduces the overhead to constant by efficiently using multi-valued Byzantine agreement (MVBA) running over a reliable broadcast which outputs a threshold signature proof that of all receivers of an input, at least one receiver is an honest peer.
SBFT~\cite{gueta2019sbft} uses threshold signatures to reduce the communication complexity to linear by extending PBFT by $c+1$ collectors.
Nodes send their messages to the collectors, who then broadcast the combined threshold signature once $3f+c+1$ shares have been received. 
Using threshold signatures reduces the message size of the collector from linear to constant, and the client overhead is reduced as only one signature has to be verified.
The execution is similarly aggregated by $c+1$ execution collectors who collect $f+1$ signature shares.
SBFT uses BLS signatures, though BLS multi-signatures instead of threshold signatures are used on the fast path as these require less computation.

Another usage of threshold cryptography is generating randomness used for leader election in asynchronous networks.
Narwhal~\cite{danezis2022narwhal} uses an adaptively secure threshold signature scheme to generate a distributed perfect coin, while Jolteon~\cite{gelashvili2021jolteon} generates randomness for each view by hashing the threshold signature of the view number in order to create an asynchronous fallback protocol to circumvent the FLP impossibility.
For ICC~\cite{camenisch2022internet}, a sequence of random beacon values is created to determine the permutation of the participants and thus the leader, similar to verifiable random functions (\cf \cref{sec:crypto:vrf}).
Starting from a known initial value, a participant generates the next sequence of the random beacon by broadcasting its signature share of the current random beacon value.
At least one honest participant's signature share is required to form a comprehensible random value.

\subsubsection{Secret Sharing}
\label{sec:crypto:secretsharing}
In secret sharing approaches, a secret is split and distributed amongst a group of $n$ nodes so that it can be reconstructed for cryptographic operations when a sufficient number (\ie a threshold $t$) of shares are combined but no single node can reconstruct the full secret by itself.
The splitting and distribution are typically performed by a dealer.
This secret sharing amongst replicas can, similar to threshold signatures, reduce the overhead of multi-signatures.
Contrary to the threshold signatures of \cref{sec:crypto:threshold}, however, the secret sharing in FastBFT~\cite{liu2018scalable} requires hardware-based trusted compartments on all replicas.
The trusted compartment on the leader can securely create secrets, split them, and deliver them to all other replicas.
This is performed in a separate pre-processing phase before the agreement phase.
Once the order has been established, the replicas release in the final step their shares of the secret to allow verification of the agreement.
All secrets are one-time secrets, and a monotonic counter value is bound to the secret shares in order to prevent equivocation of the leader.
For more details on trusted counters and trusted execution environments, we refer to \cref{sec:tee}.
Secret sharing can also be used to prevent leakage of sensitive data: clients in Qanaat~\cite{amiri2021qanaat} use $(f+1, n)$-threshold secret-sharing to keep data confidential.

\subsubsection{Erasure Coding}
\label{sec:crypto:erasurecoding}

Three protocols use erasure codes: Dumbo~\cite{guo2020dumbo}, ICC2~\cite{camenisch2022internet}, and DispersedLedger~\cite{yang2022dispersed}.
The first two aim to reduce the communication overhead and lessen the bottleneck on the leader, whereas DispersedLedger focuses on data storage and availability.
Erasure codes are forward error correction codes that can efficiently handle bit erasures during transmissions. 
In $(n, n-2t)$-erasure codes, a message $m$ is split into $n$ fragments of larger size so that any subset of $n-2t$ fragments can be used to reconstruct the original message $m$ even if fragments get lost or corrupted.
Instead of broadcasting a given payload, a leader can encode this payload and send the individual fragments to different replicas. 
With a correct encoding, the payload can be reconstructed, even if some replicas remain unresponsive. 
While reducing network load on the leader, this technique adds computational overhead for encoding and decoding.
In ICC, large blocks have to be disseminated, which can become a bottleneck, so as an improvement of their peer-to-peer gossip sub-layer (ICC0/1), they propose a subprotocol of reliable broadcast with $(n, n-2t)$-erasure codes ($t<n/3$) in ICC2 in combination with threshold signatures.
Dumbo's reliable broadcast protocol can be optimized as well using a $(n-2t, n)$-erasure code scheme in combination with a Merkle tree, which tolerates the maximal adversary boundary and thus helps honest nodes recover efficiently.
In DispersedLedger, data is stored across $n$ nodes via a verifiable information dispersal protocol making use of erasure codes.
This guarantees data availability and allows separating the tasks of agreeing on a short, ordered log and of downloading large blocks with full transactions for execution.
This decoupling of the protocol stages leads to faster progression of the protocol as the high-bandwidth task of downloading transactions is no longer on the critical path.

\subsubsection{Verifiable Random Functions}
\label{sec:crypto:vrf}

Five papers incorporate \acfp{VRF}~\cite{micali99vrf} into their design: Algorand~\cite{gilad2017algorand}, Beh-Raft-Chain~\cite{wang2021behraftchain}, Cumulus~\cite{gai2021cumulus}, Proof-of-QoS~\cite{yu2019proofofqos}, and DLattice~\cite{zhou2019dlattice}.
A VRF is a cryptographic function $\textit{VRF}_{sk}(x)$ that for an input string $x$ returns both a hash value and a proof $\pi$.
The hash value is here uniquely determined by both the user's secret key $sk$ as well as the input $x$, but appears undistinguishable from a random value for anyone not in possession of $sk$.
The proof $\pi$ allows anyone who knows the public key $pk$ to verify whether the hash value corresponds to $x$, without revealing $sk$~\cite{gilad2017algorand}.
VRFs in BFT protocols facilitate the selection of committees, which in turn can efficiently perform the consensus.
Their non-interactive nature makes them desirable as it prevents any targeted attack on the leader(s) or committee members of the next instance, as their membership status is not known in advance.
For more details on the usage of committees for scalability in BFT protocols, we refer to \cref{sec:consensus-selection}.

Algorand uses a VRF based on the nodes' key pairs as well as publicly available blockchain information for cryptographic sortition, meaning to select the committee members in a private and non-interactive way so that nodes can independently determine whether they are in the committee for this instance.
VRFs create a pseudo-random hash value that is uniformly distributed between 0 and $2^{hashlen}-1$, meaning that nodes get selected at random to be in the committee.
As VRFs do not require interaction between the nodes and are calculated using private information, a node's membership status cannot be determined in advance to launch a targeted DoS attack or allow malicious nodes to collude; instead, a node's membership status is only known retrospectively.
Algorand also includes a seed in each instance, which is calculated using the VRF result in combination with the previous instance's seed; if this seed is not included in a proposed block, it is discarded as invalid.
If multiple blocks get proposed by committee members, then the included seed value is used to determine a priority amongst these blocks.
Furthermore, the VRF value can be used as a random value for coins, for example, in order to resume normal operation after a network partition.
Algorand implements its VRFs over Curve25519~\cite{goldberg16vrfcurve, goldbe-vrf-00}, and shows that VRF and signature verification are CPU bottlenecks in the protocol.
VRFs are used similarly in Beh-Raft-Chain~\cite{wang2021behraftchain} and DLattice~\cite{zhou2019dlattice} for cryptographic sortition in order to determine a node's role and membership status, \eg (local) committee leader.
Proof-of-QoS~\cite{yu2019proofofqos} splits nodes into regions, and one node per region is selected for the BFT committee based on its quality of service~(QoS). 
Out of the $\kappa$ nodes with the highest QoS in a region, the node with the smallest hash of id and seed value is selected. 
This node's membership status is confirmed with its VRF hash, and others verify the corresponding proof. 

In Cumulus~\cite{gai2021cumulus}, VRFs are used in a novel cryptographic sortition protocol called Proof-of-Wait, which is used by this side-chain protocol to select the epoch's representative to interact with the mainchain.
At most, one representative can be chosen per epoch, and it is based on nodes calculating a random waiting time based on their VRF output.
The VRF is implemented based on the elliptic curve Secp256k1.

\subsection{Independent Groups}
In this section, we look at how subsets of all nodes, or groups, can process transactions independently from the rest of the nodes.
Of the 22 papers in this section, we classified 16 as using sharding techniques.
The remaining six papers are classified as hierarchical consensus.

Sharding, traditionally used in database systems, splits a dataset into smaller subsets.
In database systems, this is used when the data cannot fit onto a single machine anymore.
This technique has been used by \ac{BFT} protocols to improve scalability.
In a sharding algorithm, the application's state is distributed over multiple, possibly overlapping, groups called shards~\cite{hellings2020cerberus, wang2021behraftchain, hong2021pyramid, amiri2021sharper}.
This allows efficient processing of transactions within shards, as only a subset of nodes has to participate in the consensus. Further, only the responsible group must execute the transaction.

Hierarchical consensus can but does not have to split state across different groups.
Here, the groups use a higher-level consensus mechanism to coordinate.
While shards can also communicate with each other, \eg for cross-shard transactions, the higher-level consensus is a significant aspect of hierarchical consensus, which we use to distinguish these two categories.


\subsubsection{Sharding}\label{sec:sharding}


\newcommand{\threewaysubfigurewidth}{0.67\columnwidth}
\newcommand{\threewaygraphicswidth}{0.77\columnwidth}
\begin{figure*}[t]
  \centering
\begin{subfigure}[h]{\threewaysubfigurewidth}
    \centering
    \includegraphics[width=\threewaygraphicswidth]{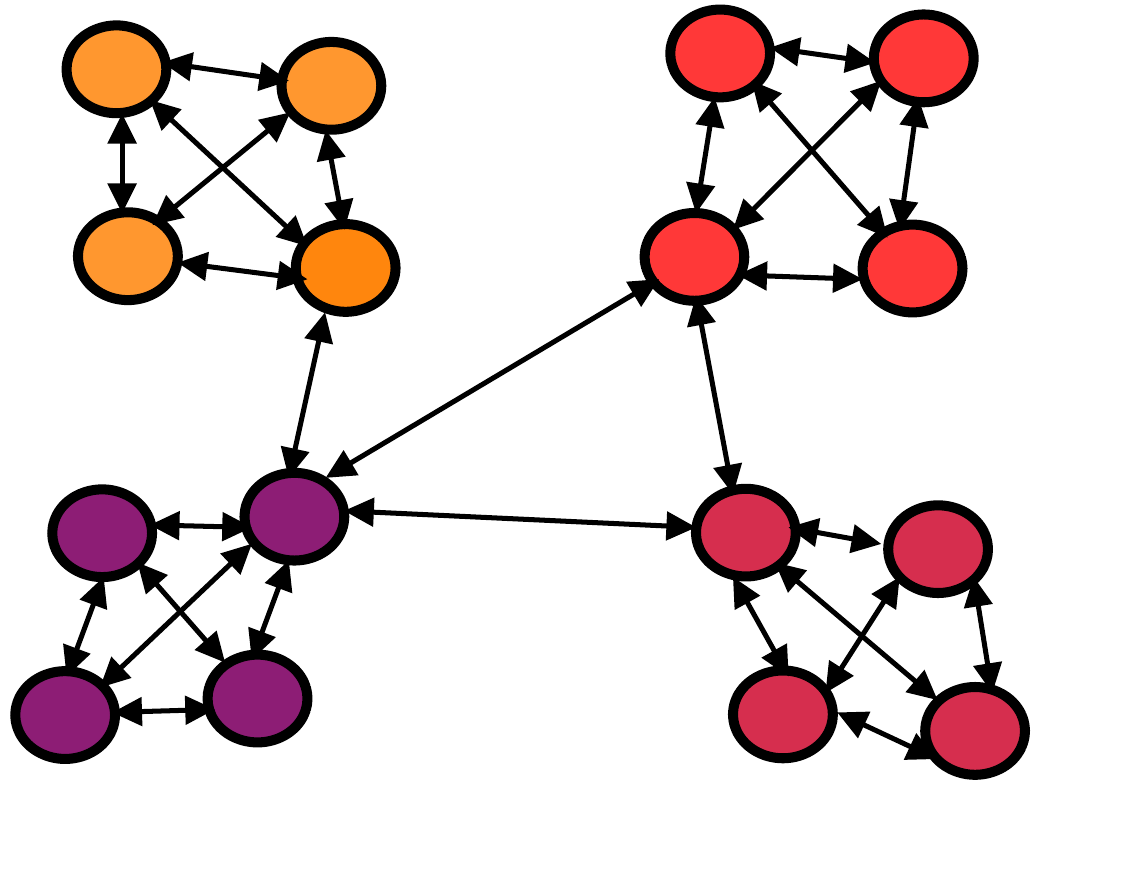}
    \caption{Replicas using sharding. Coordination between shards is commonly only required for cross-shard transactions.}
    \label{fig:sharding}
\end{subfigure}
\begin{subfigure}[h]{\threewaysubfigurewidth}
    \centering
    \includegraphics[width=\threewaygraphicswidth]{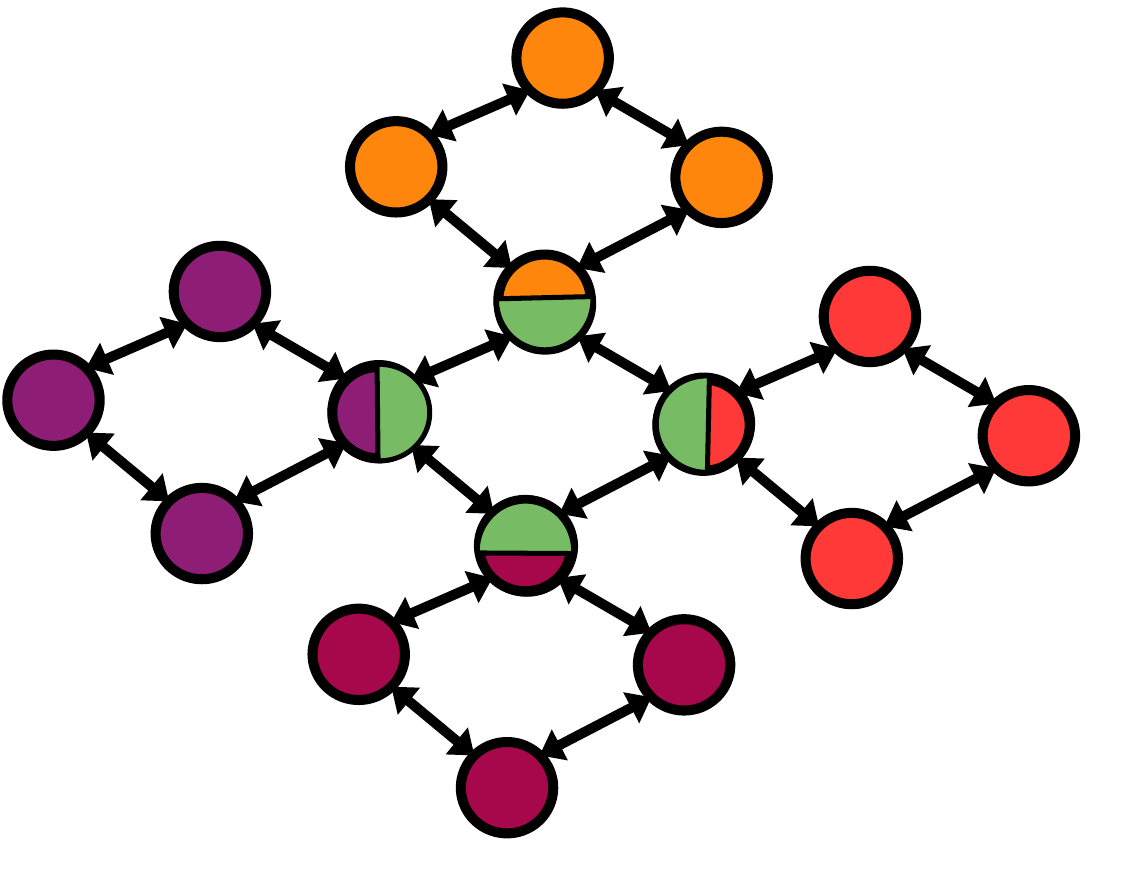}
    \caption{Hierachical consensus with two layers with ``representatives'' of the lower layer running the consensus in an upper layer.}
    \label{fig:hierarchical}
\end{subfigure}
\begin{subfigure}[h]{\threewaysubfigurewidth}
    \centering
    \includegraphics[width=0.74\columnwidth]{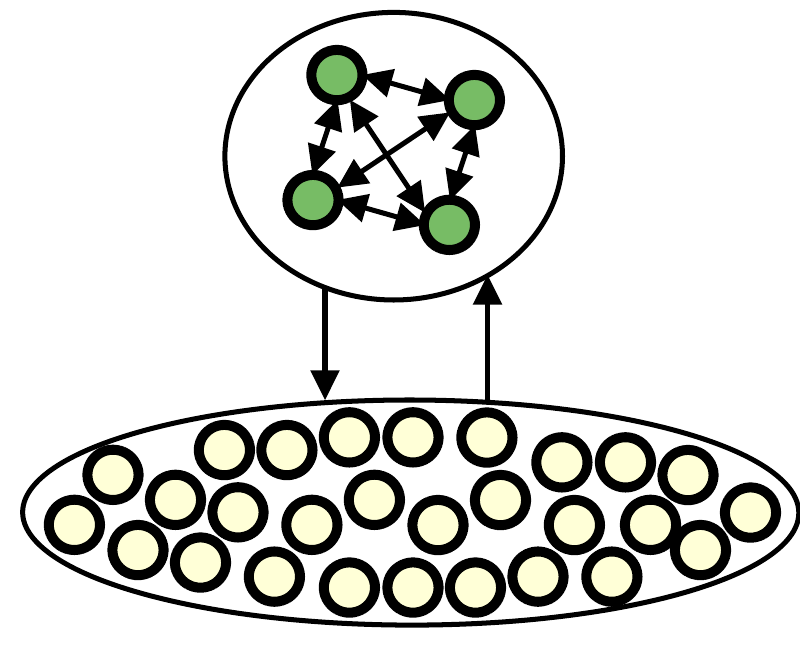}
    \caption{Schematic overview of a committee algorithm. Only a subset of all nodes performs consensus on transactions.}
    \label{fig:committee}
\end{subfigure}
\caption[]{Three ways of reducing consensus participants: sharding, hierarchical consensus, and consensus by committee.}
\label{fig:limited-consensus-participants} 
\end{figure*}

Sharding is a technique to split the system into multiple \textit{shards}, each containing a subset of the replicated state and being managed by a subset of the nodes, as seen in~\cref{fig:sharding}.
Transactions that affect only a single shard can then be processed independently. 
This improves scalability by reducing the number of messages that have to be exchanged to reach consensus within shards. 
Within a shard, nodes typically run a classic consensus protocol like PBFT~\cite{hellings2020cerberus} or Raft~\cite{wang2021behraftchain}.

Sharding also brings new challenges. 
First, as shards contain only a subset of nodes, special care must be taken so that the shards are not vulnerable.
Next, once the shards are created, clients need to know where to send transactions.
Lastly, some transactions affect multiple shards, which requires coordination between shards to process these multi-shard transactions.

The first challenge requires the protocol to assign nodes to shards while keeping the system safe and live. 
For this, different sharding algorithms assume different threat models.
Assuming a global threshold of $\frac{1}{3}$ faulty nodes, there is the risk that more than $\frac{1}{3}$ of nodes within a shard are faulty, jeopardizing safety.
This is especially a threat if an adaptive adversary is considered.
Most algorithms stipulate that the adversary can only corrupt nodes between epochs, but not within an epoch~\cite{wang2021behraftchain, hellings2020cerberus, hong2021pyramid, zamani2018rapidchain}.
This moves the challenge of safety to the shard formation.
To ensure the distribution of malicious nodes within any shard is equal to the global share of malicious nodes, techniques like \acp{VRF}~\cite{micali99vrf} or \acp{TEE}~\cite{dang2019towards} can be used.
A more challenging threat model is adaptive corruption without stipulations on when corruption can occur.
In this setting, algorithms ensure safety by resampling the shards faster than an adaptive adversary can corrupt nodes~\cite{david2021gearbox, dang2019towards}.
Some approaches form shards based on criteria such as reputation or past behavior to guarantee that well-behaved shards are formed~\cite{huang2021repchain, wang2021behraftchain}.
Chainspace~\cite{al-bassam2018chainspace} also allows for whole shards to be controlled by an adversary while still enforcing some restricted safety criteria.
Explicitly, Chainspace can still guarantee encapsulation between state objects (smart contracts) and non-repudiation in case a complete shard is controlled by the adversary. 
The encapsulation of state objects guarantees that malicious smart contracts cannot interfere with non-malicious ones, \ie the control mechanism of Chainspace.
Combined with non-repudiation, where message authorship cannot get disputed, sources of inconsistencies can get identified by the control mechanism and thus punished.



With a sharding architecture, some care has to be taken so that transactions by clients end up in the correct shard.
This is especially true for cross-shard transactions.
The simplest solution is to have clients send transactions to all shards~\cite{hellings2021byshard, hellings2020cerberus, al-bassam2018chainspace, huang2021repchain}.
This way the shards can decide if a transaction is relevant to the shard or discard the transaction otherwise.
This is more work for clients but, as ignoring transactions not relevant for a shard is easy, the overhead for the shards is small.
Transaction routing has also been adapted in different ways.
One is that clients can send transactions to an arbitrary shard, which then forwards the transaction to the correct shard~\cite{rahnama2021ringbft, amiri2021sharper}.
In the Red Belly Blockchain~\cite{crain2021red}, clients send balance and transaction requests to known proposers published in configuration blocks.
The proposers can respond directly for balance (\ie read) requests and forward transactions (\ie write requests) to consensus instances.

Lastly, transactions can affect multiple shards~\cite{hellings2021byshard, hellings2020cerberus, hong2021pyramid, amiri2021sharper, zamani2018rapidchain, amiri2021qanaat}. 
This occurs when a transaction requires data from multiple shards.
As such, these shards must coordinate access to ensure consistency.
The system then has to coordinate cross-shard transactions to make sure the transaction is accepted by each shard. 
Systems like RapidChain handle cross-shard transactions by splitting them into multiple sub-transactions that get handled by the respective shards~\cite{zamani2018rapidchain}.
This is possible as RapidChain uses the \ac{UTXO} state model, which allows splitting transactions in the spending of existing outputs and the creation of new outputs.
RapidChain calculates that more than 96\% of transactions in their system are cross-shard transactions.
As an optimization, Pyramid, which uses an account model, proposes to overlap shards, allowing cross-shard transactions to be processed by the nodes overlapping both shards~\cite{hong2021pyramid}. 
This way, overlapping nodes efficiently process cross-shard transactions, requiring no additional work. 
The overlapping nodes generate blocks from the transactions and send them to the rest of the shard for commitment.
In Cerberus~\cite{hellings2020cerberus}, each shard locally reaches consensus on \ac{UTXO} transactions.
If input from other shards is necessary, each shard will send its own local input to the other affected shards, thereby promising to process the transaction once all the required input is available by the other shards.
As all shards do the same procedure, they will either receive all required inputs and process the transaction, or all abort the transaction.
In SharPer~\cite{amiri2021sharper}, which also uses an account model, cross-shard transactions are handled by clients sending the transactions to some leader of a shard from which data is needed.
This leader is then responsible for sending the transaction to all nodes of all other involved shards.

\subsubsection{Hierarchical Consensus}
\label{subsec:independent_hierachical}

In a hierarchical consensus scheme, nodes on lower hierarchy levels can operate in independent groups and use higher levels to coordinate.
The papers in this section use different approaches to improve scalability.
The biggest benefit, as with sharding, is that the smaller groups allow for more efficient communication, as the number of communicating nodes is reduced~\cite{xu2021concurrent, amiri2021saguaro, zheng2020dphybrid}.

This architecture poses multiple questions.
Firstly, how is the hierarchy structured, and how is it determined?
How many layers are there, and what information has to get shared between layers?
Finally, and most importantly for scalability, what degree of independence can lower levels have, and when are higher levels of consensus necessary?

Regarding the hierarchy structure, of the papers considered, four picked a two-layer hierarchy~\cite{zheng2020dphybrid, li2021optimized, wang2021behraftchain, xu2021concurrent}, while two chose a variable number of layers~\cite{li2021scalable, amiri2021saguaro}.
For the two-layer structure, a structure using ``representatives'' is most common~\cite{li2021optimized, wang2021behraftchain, xu2021concurrent}, which is also used by X-Layer~\cite{li2021scalable} for multiple layers.
In this case, as seen in \cref{fig:hierarchical}, the lower levels have their ``representative'', commonly leaders, representing the group in the upper layer.
Similarly, GeoBFT~\cite{gupta2020resilientdb} shares transactions globally by sending them to the leaders of other clusters before executing them.
The difference is that global sharing does not require coordination between the leaders, but the leaders share the transaction within their own cluster.
In another way, DP-Hybrid~\cite{zheng2020dphybrid} chooses to use \ac{PoW} as the mechanism for the upper layer where all nodes can participate.
Focusing on processing transactions in wide area networks, Saguaro~\cite{amiri2021saguaro} uses multiple layers, with different layers composed of edge devices, edge servers, fog servers, and cloud servers.
The structure of hierarchies can be dynamic~\cite{wang2021behraftchain, li2021optimized}, where new shards are created if enough new nodes join the network, or static as in the cases of DP-Hybrid~\cite{zheng2020dphybrid} and Saguaro~\cite{amiri2021saguaro}, where the algorithm is focused on companies using the blockchain.

Next is the question on what information the higher level works.
There are two broad approaches to this: either with the actual transactions~\cite{zheng2020dphybrid, gupta2020resilientdb, li2021optimized, amiri2019caper}, or with ``prepackaged'' blocks of transactions~\cite{xu2021concurrent, amiri2021saguaro} arranged in the upper layer.
For example, in GeoBFT~\cite{gupta2020resilientdb} or DP-Hybrid~\cite{zheng2020dphybrid}, every transaction is shared through the upper layer with the other groups.
In C-PBFT~\cite{xu2021concurrent}, on the other hand, the lower layer already constructs blocks of transactions, which are then confirmed by adding the block header to the upper-layer blockchain.

As with sharding, it can happen that some transaction requires data from multiple groups.
For this, affected groups can coordinate with their least common ancestor group~\cite{amiri2021saguaro}.
As all transactions are globally shared in GeoBFT~\cite{gupta2020resilientdb}, each cluster has access to all transactions, making cross-cluster transactions cheap.
As global sharing is performed before execution, this leads to higher latencies.

Different layers can have different failure modes~\cite{amiri2021saguaro, amiri2019caper}, \eg when different applications use a shared blockchain.
This can be used to optimize the system, \eg if the applications require different security or fault tolerance levels.
Then it might be possible to use different consensus protocols in separate lower layers, such as BFT or crash fault-tolerant protocols.

Further, there are protocols that measure the behavior of nodes and reward good behavior, \eg to allow well-acting nodes to become leaders more often~\cite{wang2021behraftchain, xu2021concurrent}.
The behavior-measuring metrics consist of the detection of misbehaving nodes~\cite{wang2021behraftchain} and qualitative measures like payment times and amounts~\cite{xu2021concurrent}.
\enquote{Good} participants are thus rewarded for their good metrics, creating incentives for good behavior.

\subsection{Selection of Consensus Committees}\label{sec:consensus-selection}
As described in~\cref{sec:communication-topologies}, one approach to improve scalability is to avoid bottleneck situations.
In this direction, several approaches regarding nodes' roles for agreement exist.
Of the 20 papers in this section, 10 focus on consensus by committee, eight on hierarchical consensus (\cf \cref{subsec:independent_hierachical}), and two on randomized sampling.
One approach is that only a substantially smaller subset of participants, \ie committees, participate in finding agreement~\cite{gilad2017algorand, zamani2018rapidchain, jalalzai2019proteus, jalalzai2021hermes}.
As these committees are significantly smaller than the whole set of participants, consensus protocols---which, in theory, would not scale for the whole set of nodes---become sufficiently efficient again.
The remaining nodes often take a passive role and only observe and verify the committee's work.
A related technique is a hierarchical consensus where multiple layers are used for consensus~\cite{li2021scalable, li2021optimized, amiri2019caper, xu2021concurrent, amiri2021saguaro}, which faces similar problems as consensus by committee.
In this section, we investigate protocols using these techniques to see how committees are formed, how they work, and what aspects should be considered for scalability.

\subsubsection{Committee}
In a committee, only a subset of nodes participates in the consensus algorithm, as depicted in~\cref{fig:committee}, while non-committee members observe the results of the consensus~\cite{gilad2017algorand, zhan2021drbft, jalalzai2021hermes}.
As only a substantially smaller subset of the total nodes participate in the consensus algorithm, the communication efforts needed inside of the committee are reduced.
The crucial question is: how is it decided which participants will be included in the committee without being vulnerable to attacks?
The main committee formation scheme is based on randomness.
Randomness makes it unlikely that a critical amount of malicious nodes will enter the committee.
Different algorithms use different randomness to select committee nodes.
For example, Algorand utilizes \acp{VRF}~\cite{gilad2017algorand} (\cf \cref{sec:crypto:vrf}), DRBFT picks committee nodes based on previous blocks~\cite{zhan2021drbft}, and RapidChain uses verifiable secret sharing to generate randomness within committees~\cite{zamani2018rapidchain}.
Some algorithms modify the committee formation based on metrics such as stake~\cite{gilad2017algorand, zhou2019dlattice} or quality characteristics~\cite{yu2019proofofqos} such as bandwidth or latency to prioritize nodes by some weight.
In permissionless protocols like Algorand, the stake is necessary to prevent Sybil attacks based on pseudonyms.
Otherwise, attackers could create an arbitrary number of nodes to gain control of the committee to influence the consensus maliciously.
The weighting of quality metrics incentivizes nodes to provide more efficient services to become committee nodes.
Thus, preferred nodes are presumably better committee members than randomly selected ones.

Malicious committee members pose a risk to the safety and liveness of protocols.
Inside the committee, \ac{BFT} protocols are used to tolerate malicious behavior.
Additionally, some algorithms use different mechanisms to reduce the risk of a malicious committee.
Commonly the results of the committee are verified.
This verification is both to check that progress is made, \eg that blocks are proposed, but also of the blocks themselves~\cite{zhan2021drbft, jalalzai2019proteus}.
For example, in Proteus, a committee is replaced with new members if the committee does not generate valid blocks~\cite{jalalzai2019proteus}.
To encourage participation, committee members who act maliciously risk forfeiture of their stake in protocols requiring a deposit to join the committee~\cite{zhou2019dlattice}.


By definition, committees consist of only a subset of the total nodes.
This opens up the risk that an adversary could gain control of a committee by chance or by adaptive corruptions while controlling less than $\frac{1}{3}$ of the total nodes.
One way to overcome this risk of adaptive corruption is to change out the committee members regularly~\cite{matt2022formalizing}.
For example, in Algorand~\cite{gilad2017algorand} any committee member is replaced after any message sent by the member.
Thus, committee members are immediately replaced once identified as potential victims, making it impossible for attackers to target committee members.
In contrast, for Dumbo~\cite{guo2020dumbo} or in PoQ~\cite{yu2019proofofqos}, committees are persistent for one block generation.
This makes committee members vulnerable but limits the attack impact to one block.
Similarly, in AHL each committee is replaced after every epoch~\cite{dang2019towards}. 
This is assumed to be safe even in the presence of an adaptive attacker model where nodes are not instantly corrupted but rather after some time.
RapidChain~\cite{zamani2018rapidchain} also replaces committee members after every epoch, though as an optimization, it replaces only a subset of committee members.
AHL and RapidChain make limitations on the threat model.
AHL relies on TEEs, which are assumed to only fail by crashing. 
RapidChain assumes that the adaptive adversary can only corrupt nodes at the start of the protocol and in between epochs, but not within an epoch.
In Hermes~\cite{jalalzai2021hermes}, which does not assume an adaptive adversary, committees can get replaced by view changes.
Non-committee members initiate a view change after not receiving a block over some period of time, after receiving an invalid block proposal, or after multiple proposals with the same sequence number.

\subsubsection{Randomized Sampling}
\label{randomized-sampling}
We have seen how leaders can become the bottleneck regarding scalability.
Randomized sampling can be used to create leaderless consensus protocols, thus avoiding bottlenecks at a single leader~\cite{rocket2020avalanche, lim2014scalable}.
With randomized sampling, nodes only communicate with a subset of the total nodes.
In the process, the nodes exchange information locally about their values.
These values can be the value a node has locally decided as in~\cite{rocket2020avalanche}, or DecicionVectors as in~\cite{lim2014scalable} where the decisions of all nodes are exchanged.
A global consensus is achieved by running these local updates multiple times so that, over time the nodes converge to a single decision.
With these local updates, there is no need for a leader to drive the consensus forward.
This removes the common bottleneck in other consensus protocols, as no single node is responsible for disseminating or collecting any information to and from all other nodes.
The required communication stays constant for each node as it only needs to exchange information with a configurable but constant subset of local nodes~\cite{rocket2020avalanche, lim2014scalable}.
Nodes learn of proposed values from others.
In Avalanche, this happens after the node queries neighboring nodes, while in ~\cite{lim2014scalable}, nodes can also actively push their value to other nodes.
However, they only provide probabilistic safety guarantees. 



\subsection{Hardware Support}
\label{sec:tee}





The scalability of \ac{BFT} protocols can also be improved via hardware support as offered by \acfp{TEE}, as has been done in two papers matching our criteria: FastBFT~\cite{liu2018scalable} and AHL~\cite{dang2019towards}.
The most commonly used \ac{TEE} is \ac{SGX} due to its high availability; however, many approaches are independent of the underlying \ac{TEE}, allowing the use of \eg \acp{TPM}, ARM TrustZone, or AMD SEV-SNP.
\ac{SGX} is an x86 instruction set extension that allows the creation of so-called \emph{enclaves}, in which confidentiality of execution and integrity of data is ensured from privileged software via hardware-based memory encryption and checksums.
Enclaves can be remotely attested: users can verify an enclave's code and data, and before execution of the enclave, the provided code and data is hashed to create a measurement.
This measurement can then be compared against the value of the verified enclave to ensure that the user communicates with a genuine, correct, and unmodified version of the expected enclave. 

\acp{TEE} can therefore be used as a trusted subsystem in \ac{BFT} protocols with a \emph{hybrid} fault model:
while the remainder of the \ac{BFT} system can still behave arbitrarily faulty, the trusted subsystem is assumed to behave correctly and can only fail by crashing.
This can be used to prevent equivocation, \ie sending conflicting messages to different communication partners in the protocol.
Primitives that make use of such trusted subsystems are \eg trusted counters~\cite{levin09trinc} or attested append-only memory~\cite{chun07a2m}.
FastBFT~\cite{liu2018scalable} employs trusted monotonic counters that are provided by the \ac{TEE} running on the leader replica.
A counter value extends every message sent by the leader, and as every value can only be used once and the counter is monotonically increasing, it can thus be detected if the leader equivocates.
Relying on trusted hardware and therefore preventing equivocation allows reducing the complexity of \ac{BFT} protocols, \eg by decreasing the number of replicas from $3f+1$ to $2f+1$ or the required communication rounds for agreement, or by using less expensive cryptographic primitives (see also \cref{sec:crypto}). 

\acp{TEE} are also used to efficiently aggregate messages \cite{dang2019towards, liu2018scalable}, \eg by combining a quorum of $2f+1$ messages into a proof issued by the \ac{TEE}~\cite{dang2019towards}.
Here, the leader collects and aggregates other nodes' signatures into a single authenticated message, while nodes forward their signed messages to the leader and verify the created multi-signature.
%
Furthermore, \acp{TEE} can also be used as a source for a trusted randomness beacon, which can be used to efficiently partition the system for sharding~\cite{dang2019towards} (see also \cref{sec:sharding}).

\newcommand{\cell}[2]{\multicolumn{1}{c|}{\cellcolor[HTML]{#1}{#2}}}
\newcommand{\doublecell}[2]{\begin{tabular}[c]{@{}c@{}}{#1}\\ {#2}\end{tabular}}

\newcommand{\conf}{conf.}

\newcommand{\networkunk}{network?}
\newcommand{\networksync}{sync}
\newcommand{\networkasync}{async}
\newcommand{\networkpartialsync}{part. sync}

\newcommand{\maxfhalf}{$f<\frac{1}{2}n$}
\newcommand{\maxfthird}{$f<\frac{1}{3}n$}
\newcommand{\maxfonethreen}{1/3n}
\newcommand{\maxfunk}{maxf?}
\newcommand{\maxfpowbound}{PoW bound}
\newcommand{\maxfparameterizable}{paramet.}
\newcommand{\maxfresdb}{\doublecell{$f<\frac{1}{3}n$}{per cluster}}
\newcommand{\maxfSBFT}{\doublecell{$n=$}{$3f+2c+1$}}

\newcommand{\membershipdynamic}{dynamic}
\newcommand{\membershipstatic}{static}

\newcommand{\safetydeterministic}{det.}
\newcommand{\safetyprobabilistic}{prob.}

\newcommand{\leadernone}{none}
\newcommand{\leadermulti}{multi}
\newcommand{\leaderpowelected}{PoW-elected}
\newcommand{\leaderrandomselection}{random select.}
\newcommand{\leaderrotating}{rotating}
\newcommand{\leadersingle}{single}

\newcommand{\cryptovrf}{VRF}
\newcommand{\cryptoaggregate}{aggregate}
\newcommand{\cryptoaggregation}{\cryptoaggregate}
\newcommand{\cryptobls}{BLS}
\newcommand{\cryptoerasurecoding}{erasure coding}
\newcommand{\cryptothreshold}{threshold sig.}
\newcommand{\cryptomultisignature}{multi sig.}
\newcommand{\cryptosecretsharing}{secret sharing}
\newcommand{\cryptothresholdencryption}{\cryptothreshold}

\newcommand{\exchangegossip}{gossip}
\newcommand{\exchangestar}{star}
\newcommand{\exchangeclique}{clique}
\newcommand{\exchangering}{ring}
\newcommand{\exchangetree}{tree}

\newcommand{\pipelininghotstuff}{c.f. HotStuff}

\newcommand{\parallelizationsharding}{sharding}
\newcommand{\parallelizationclustering}{clustering}
\newcommand{\parallelizationhierachicalconsensus}{hierarchical}

\newcommand{\consensushierarchical}{hierarchical}
\newcommand{\consensuscommittee}{committee}
\newcommand{\consensusrandomizedsampling}{rand. sampling}
\newcommand{\consensusspeculativeexecution}{\doublecell{speculative}{execution}}

\newcommand{\colorassumption}{C9FEFC}
\newcommand{\colorguarantees}{E0FFE0}
\newcommand{\colorscaling}{FFFFC7}

\begin{table*}[!ht]
\centering
\caption{Comparison of scalable \ac{BFT} protocols regarding their assumptions, goals, and scaling techniques.}
\label{tab:protocols}
\resizebox{\textwidth}{!}{%
\begin{tabular}{|c|
>{\columncolor[HTML]{C9FEFC}}c 
>{\columncolor[HTML]{C9FEFC}}c 
>{\columncolor[HTML]{C9FEFC}}c |
>{\columncolor[HTML]{E0FFE0}}c 
>{\columncolor[HTML]{E0FFE0}}c |
>{\columncolor[HTML]{FFFFC7}}c 
>{\columncolor[HTML]{FFFFC7}}c 
>{\columncolor[HTML]{FFFFC7}}c 
>{\columncolor[HTML]{FFFFC7}}c 
>{\columncolor[HTML]{FFFFC7}}c 
>{\columncolor[HTML]{FFFFC7}}c 
>{\columncolor[HTML]{FFFFC7}}c |}
\hline
\cellcolor[HTML]{FFFFFF}{\color[HTML]{000000} } &
  \multicolumn{3}{c|}{\cellcolor[HTML]{\colorassumption}{\color[HTML]{000000} \textbf{Assumptions}}} &
  \multicolumn{2}{c|}{\cellcolor[HTML]{\colorguarantees}{\color[HTML]{000000} \textbf{Guarantees}}} &
  \multicolumn{7}{c|}{\cellcolor[HTML]{\colorscaling}{\color[HTML]{000000} \textbf{Scaling Techniques}}} \\ \cline{2-13} 
\multirow{-2}{*}{\cellcolor[HTML]{FFFFFF}{\color[HTML]{000000} \textbf{\begin{tabular}[c]{@{}c@{}}BFT\\ Protocol\end{tabular}}}} &
  \multicolumn{1}{c|}{\cellcolor[HTML]{\colorassumption}{\color[HTML]{000000} \textbf{\begin{tabular}[c]{@{}c@{}}Network (for\\ safety+liveness)\end{tabular} }}} &
  \multicolumn{1}{c|}{\cellcolor[HTML]{\colorassumption}{\color[HTML]{000000} \textbf{Fault}}} &
  {\color[HTML]{000000} \textbf{Members}} &
  \multicolumn{1}{c|}{\cellcolor[HTML]{\colorguarantees}{\color[HTML]{000000} \textbf{Safety}}} &
  {\color[HTML]{000000} \textbf{Liveness}} &
  \multicolumn{1}{c|}{\cellcolor[HTML]{\colorscaling}{\color[HTML]{000000} \textbf{Leader}}} &
  \multicolumn{1}{c|}{\cellcolor[HTML]{\colorscaling}{\color[HTML]{000000} \textbf{Crypto}}} &
  \multicolumn{1}{c|}{\cellcolor[HTML]{\colorscaling}{\color[HTML]{000000} \textbf{\begin{tabular}[c]{@{}c@{}}Message\\ Exchange\end{tabular}}}} &
  \multicolumn{1}{c|}{\cellcolor[HTML]{\colorscaling}{\color[HTML]{000000} \textbf{\begin{tabular}[c]{@{}c@{}}Pipelining\\ Strategy\end{tabular}}}} &
  \multicolumn{1}{c|}{\cellcolor[HTML]{\colorscaling}{\color[HTML]{000000} \textbf{Consensus}}} &
  \multicolumn{1}{c|}{\cellcolor[HTML]{\colorscaling}{\color[HTML]{000000} \textbf{\begin{tabular}[c]{@{}c@{}}Parallel-\\ization\end{tabular}}}} &
  \textbf{Other} \\ \hline

ByzCoin \cite{kogias2016enhancing} 
&
\cell{\colorassumption}{\networkpartialsync} & 
\cell{\colorassumption}{\maxfpowbound} & 
\cell{\colorassumption}{\membershipdynamic} & 
\cell{\colorguarantees}{\safetydeterministic} & 
\cell{\colorguarantees}{\safetydeterministic} & 
\cell{\colorscaling}{\doublecell{\leadersingle}{\leaderpowelected}} & 
\cell{\colorscaling}{\cryptomultisignature} & 
\cell{\colorscaling}{\exchangetree} & 
\cell{\colorscaling}{-} & 
\cell{\colorscaling}{-} & 
\cell{\colorscaling}{-} & 
\cell{\colorscaling}{-}  
 \\ \hline

Lim et al.~\cite{lim2014scalable} &
\cell{\colorassumption}{\networkasync} & 
\cell{\colorassumption}{\maxfhalf} & 
\cell{\colorassumption}{\membershipstatic} & 
\cell{\colorguarantees}{\safetyprobabilistic} & 
\cell{\colorguarantees}{\safetyprobabilistic} & 
\cell{\colorscaling}{\leadernone} & 
\cell{\colorscaling}{-} & 
\cell{\colorscaling}{\exchangegossip} & 
\cell{\colorscaling}{-} & 
\cell{\colorscaling}{\consensusrandomizedsampling} & 
\cell{\colorscaling}{-} & 
\cell{\colorscaling}{-}  
 \\ \hline
 
Tendermint \cite{buchman2016tendermint} &
\cell{\colorassumption}{\networkpartialsync} & 
\cell{\colorassumption}{\maxfthird} & 
\cell{\colorassumption}{\membershipdynamic} & 
\cell{\colorguarantees}{\safetydeterministic} & 
\cell{\colorguarantees}{\safetyprobabilistic} & 
\cell{\colorscaling}{\doublecell{\leadersingle}{\leaderrotating}} & 
\cell{\colorscaling}{-} & 
\cell{\colorscaling}{\exchangegossip} & 
\cell{\colorscaling}{-} & 
\cell{\colorscaling}{-} & 
\cell{\colorscaling}{-} & 
\cell{\colorscaling}{-}  
 \\ \hline

RepChain \cite{huang2021repchain} &
\cell{\colorassumption}{\networksync} & 
\cell{\colorassumption}{\maxfthird} & 
\cell{\colorassumption}{\membershipdynamic} & 
\cell{\colorguarantees}{\safetyprobabilistic} & 
\cell{\colorguarantees}{\safetyprobabilistic} & 
\cell{\colorscaling}{\leadersingle} & 
\cell{\colorscaling}{\cryptomultisignature} & 
\cell{\colorscaling}{-} & 
\cell{\colorscaling}{-} & 
\cell{\colorscaling}{-} & 
\cell{\colorscaling}{\parallelizationsharding} & 
\cell{\colorscaling}{-}  
 \\ \hline

Ostraka \cite{manuskin2020ostraka} &
\cell{\colorassumption}{\conf} & 
\cell{\colorassumption}{\maxfhalf} & 
\cell{\colorassumption}{\membershipdynamic} & 
\cell{\colorguarantees}{\conf} & 
\cell{\colorguarantees}{\conf} & 
\cell{\colorscaling}{\conf} & 
\cell{\colorscaling}{-} & 
\cell{\colorscaling}{-} & 
\cell{\colorscaling}{-} & 
\cell{\colorscaling}{-} & 
\cell{\colorscaling}{\parallelizationsharding} & 
\cell{\colorscaling}{-}  
 \\ \hline

Mitosis \cite{marson2021mitosis} &
\cell{\colorassumption}{\networkpartialsync} & 
\cell{\colorassumption}{\maxfparameterizable} & 
\cell{\colorassumption}{\membershipdynamic} & 
\cell{\colorguarantees}{\conf} & 
\cell{\colorguarantees}{\conf} & 
\cell{\colorscaling}{\conf} & 
\cell{\colorscaling}{-} & 
\cell{\colorscaling}{-} & 
\cell{\colorscaling}{-} & 
\cell{\colorscaling}{-} & 
\cell{\colorscaling}{\parallelizationsharding} & 
\cell{\colorscaling}{-}  
 \\ \hline

Kauri \cite{neiheiser2021kauri} &
\cell{\colorassumption}{\networkpartialsync} & 
\cell{\colorassumption}{\maxfthird} & 
\cell{\colorassumption}{\membershipstatic} & 
\cell{\colorguarantees}{\safetydeterministic} & 
\cell{\colorguarantees}{\safetydeterministic} & 
\cell{\colorscaling}{\leadersingle} & 
\cell{\colorscaling}{\cryptomultisignature} & 
\cell{\colorscaling}{\exchangetree} & 
\cell{\colorscaling}{\doublecell{configurable pipe-} {lining stretch}} & 
\cell{\colorscaling}{-} & 
\cell{\colorscaling}{-} & 
\cell{\colorscaling}{-}  
 \\ \hline

HotStuff \cite{hotstuff19} &
\cell{\colorassumption}{\networkpartialsync} & 
\cell{\colorassumption}{\maxfthird} & 
\cell{\colorassumption}{\membershipstatic} & 
\cell{\colorguarantees}{\safetydeterministic} & 
\cell{\colorguarantees}{\safetydeterministic} & 
\cell{\colorscaling}{\doublecell{\leadersingle}{\leaderrotating}} & 
\cell{\colorscaling}{\cryptothreshold} & 
\cell{\colorscaling}{\exchangestar} & 
\cell{\colorscaling}{\doublecell{one pipeline stage}{per protocol stage}} & 
\cell{\colorscaling}{-} & 
\cell{\colorscaling}{-} & 
\cell{\colorscaling}{-}  
 \\ \hline

X-Layer PBFT \cite{li2021scalable} &
\cell{\colorassumption}{\networkpartialsync} & 
\cell{\colorassumption}{\maxfparameterizable} & 
\cell{\colorassumption}{\membershipstatic} & 
\cell{\colorguarantees}{\safetyprobabilistic} & 
\cell{\colorguarantees}{\safetyprobabilistic} & 
\cell{\colorscaling}{\leadersingle} & 
\cell{\colorscaling}{-} & 
\cell{\colorscaling}{\exchangetree} & 
\cell{\colorscaling}{-} & 
\cell{\colorscaling}{\consensushierarchical} & 
\cell{\colorscaling}{-} & 
\cell{\colorscaling}{-}  
 \\ \hline


CHECO \cite{cong2018blockchain} &
\cell{\colorassumption}{\networkasync} & 
\cell{\colorassumption}{\maxfthird} & 
\cell{\colorassumption}{\membershipdynamic} & 
\cell{\colorguarantees}{\safetyprobabilistic} & 
\cell{\colorguarantees}{\safetyprobabilistic} & 
\cell{\colorscaling}{\leadermulti} & 
\cell{\colorscaling}{-} & 
\cell{\colorscaling}{-} & 
\cell{\colorscaling}{-} & 
\cell{\colorscaling}{-} & 
\cell{\colorscaling}{-} & 
\cell{\colorscaling}{-}  
 \\ \hline

Beh-Raft-Chain \cite{wang2021behraftchain} &
\cell{\colorassumption}{\networksync} & 
\cell{\colorassumption}{\maxfthird} & 
\cell{\colorassumption}{\membershipdynamic} & 
\cell{\colorguarantees}{\safetyprobabilistic} & 
\cell{\colorguarantees}{\safetyprobabilistic} & 
\cell{\colorscaling}{\leadersingle} & 
\cell{\colorscaling}{\cryptovrf} & 
\cell{\colorscaling}{-} & 
\cell{\colorscaling}{-} & 
\cell{\colorscaling}{\consensushierarchical} & 
\cell{\colorscaling}{\parallelizationsharding} & 
\cell{\colorscaling}{-}  
 \\ \hline

SHBFT \cite{li2021optimized}&
\cell{\colorassumption}{\networkpartialsync} & 
\cell{\colorassumption}{\maxfthird} & 
\cell{\colorassumption}{\membershipdynamic} & 
\cell{\colorguarantees}{\safetydeterministic} & 
\cell{\colorguarantees}{\safetydeterministic} & 
\cell{\colorscaling}{\doublecell{\leadersingle}{\leadermulti}} & 
\cell{\colorscaling}{-} & 
\cell{\colorscaling}{-} & 
\cell{\colorscaling}{-} & 
\cell{\colorscaling}{\consensushierarchical} & 
\cell{\colorscaling}{\parallelizationhierachicalconsensus} & 
\cell{\colorscaling}{-}  
 \\ \hline

Hermes \cite{jalalzai2021hermes} &
\cell{\colorassumption}{\networkpartialsync} & 
\cell{\colorassumption}{\maxfthird} & 
\cell{\colorassumption}{\membershipdynamic} & 
\cell{\colorguarantees}{\safetyprobabilistic} & 
\cell{\colorguarantees}{\safetyprobabilistic} & 
\cell{\colorscaling}{\leadersingle} & 
\cell{\colorscaling}{-} & 
\cell{\colorscaling}{-} & 
\cell{\colorscaling}{\doublecell{chain-based}{pipelining}} & 
\cell{\colorscaling}{\consensuscommittee} & 
\cell{\colorscaling}{-} & 
\cell{\colorscaling}{-}  
 \\ \hline




SharPer \cite{amiri2021sharper} &
\cell{\colorassumption}{\networkpartialsync} & 
\cell{\colorassumption}{\maxfthird} & 
\cell{\colorassumption}{\membershipdynamic} & 
\cell{\colorguarantees}{\safetydeterministic} & 
\cell{\colorguarantees}{\safetydeterministic} & 
\cell{\colorscaling}{\leadersingle} & 
\cell{\colorscaling}{-} & 
\cell{\colorscaling}{-} & 
\cell{\colorscaling}{-} & 
\cell{\colorscaling}{-} & 
\cell{\colorscaling}{\parallelizationsharding} & 
\cell{\colorscaling}{-}  
 \\ \hline

FastBFT \cite{liu2018scalable} &
\cell{\colorassumption}{\networkpartialsync} & 
\cell{\colorassumption}{\maxfhalf} & 
\cell{\colorassumption}{\membershipdynamic} & 
\cell{\colorguarantees}{\safetydeterministic} & 
\cell{\colorguarantees}{\safetydeterministic} & 
\cell{\colorscaling}{\leadersingle} & 
\cell{\colorscaling}{\cryptosecretsharing} & 
\cell{\colorscaling}{\exchangetree} & 
\cell{\colorscaling}{-} & 
\cell{\colorscaling}{-} & 
\cell{\colorscaling}{-} & 
\cell{\colorscaling}{HW/TEE}  
 \\ \hline

ResilientDB \cite{gupta2020resilientdb} &
\cell{\colorassumption}{\networksync} & 
\cell{\colorassumption}{\maxfresdb} & 
\cell{\colorassumption}{\membershipstatic} & 
\cell{\colorguarantees}{\safetydeterministic} & 
\cell{\colorguarantees}{\safetydeterministic} & 
\cell{\colorscaling}{\leadermulti} & 
\cell{\colorscaling}{-} & 
\cell{\colorscaling}{-} & 
\cell{\colorscaling}{c.f. PoE} & 
\cell{\colorscaling}{-} & 
\cell{\colorscaling}{\parallelizationclustering} & 
\cell{\colorscaling}{-}  
 \\ \hline
 
BFT-Store~\cite{qi2020reliable} &
\cell{\colorassumption}{\networkpartialsync} & 
\cell{\colorassumption}{\maxfparameterizable} & 
\cell{\colorassumption}{\membershipdynamic} & 
\cell{\colorguarantees}{\safetydeterministic} & 
\cell{\colorguarantees}{\safetydeterministic} & 
\cell{\colorscaling}{\doublecell{\leadersingle}{\leaderrotating}} & 
\cell{\colorscaling}{\cryptoerasurecoding} & 
\cell{\colorscaling}{-} & 
\cell{\colorscaling}{-} & 
\cell{\colorscaling}{-} & 
\cell{\colorscaling}{-} & 
\cell{\colorscaling}{\doublecell{storage}{scalability}}  
 \\ \hline
 

Red Belly~\cite{crain2021red} &
\cell{\colorassumption}{\networkpartialsync} & 
\cell{\colorassumption}{\maxfthird} & 
\cell{\colorassumption}{\membershipdynamic} & 
\cell{\colorguarantees}{\safetydeterministic} & 
\cell{\colorguarantees}{\safetydeterministic} & 
\cell{\colorscaling}{\leadernone} & 
\cell{\colorscaling}{-} & 
\cell{\colorscaling}{-} & 
\cell{\colorscaling}{-} & 
\cell{\colorscaling}{-} & 
\cell{\colorscaling}{\parallelizationsharding} & 
\cell{\colorscaling}{-}  
 \\ \hline

RCC \cite{rcc2021gupta} &
\cell{\colorassumption}{\networkpartialsync} & 
\cell{\colorassumption}{\maxfthird} & 
\cell{\colorassumption}{\membershipstatic} & 
\cell{\colorguarantees}{\safetydeterministic} & 
\cell{\colorguarantees}{\safetydeterministic} & 
\cell{\colorscaling}{\leadermulti} & 
\cell{\colorscaling}{-} & 
\cell{\colorscaling}{-} & 
\cell{\colorscaling}{\doublecell{pipelining blocks}{across rounds}} & 
\cell{\colorscaling}{-} & 
\cell{\colorscaling}{-} & 
\cell{\colorscaling}{-}  
 \\ \hline

RapidChain~\cite{zamani2018rapidchain} &
\cell{\colorassumption}{\networksync} & 
\cell{\colorassumption}{\maxfthird} & 
\cell{\colorassumption}{\membershipdynamic} & 
\cell{\colorguarantees}{\safetyprobabilistic} & 
\cell{\colorguarantees}{\safetyprobabilistic} & 
\cell{\colorscaling}{\doublecell{\leadersingle}{\leaderrandomselection}} & 
\cell{\colorscaling}{-} & 
\cell{\colorscaling}{\exchangegossip} & 
\cell{\colorscaling}{\doublecell{intrashard}{pipelining}} & 
\cell{\colorscaling}{\consensuscommittee} & 
\cell{\colorscaling}{\parallelizationsharding} & 
\cell{\colorscaling}{-}  
 \\ \hline

Pyramid~\cite{hong2021pyramid} &
\cell{\colorassumption}{\networkpartialsync} & 
\cell{\colorassumption}{\maxfthird} & 
\cell{\colorassumption}{\membershipdynamic} & 
\cell{\colorguarantees}{\safetyprobabilistic} & 
\cell{\colorguarantees}{\safetyprobabilistic} & 
\cell{\colorscaling}{\leadermulti} & 
\cell{\colorscaling}{-} & 
\cell{\colorscaling}{-} & 
\cell{\colorscaling}{-} & 
\cell{\colorscaling}{-} & 
\cell{\colorscaling}{\parallelizationsharding} & 
\cell{\colorscaling}{-}  
 \\ \hline

Proteus~\cite{jalalzai2019proteus} &
\cell{\colorassumption}{\networkasync} & 
\cell{\colorassumption}{\maxfthird} & 
\cell{\colorassumption}{\membershipstatic} & 
\cell{\colorguarantees}{\safetyprobabilistic} & 
\cell{\colorguarantees}{\safetyprobabilistic} & 
\cell{\colorscaling}{\leadersingle} & 
\cell{\colorscaling}{-} & 
\cell{\colorscaling}{\exchangestar} & 
\cell{\colorscaling}{-} & 
\cell{\colorscaling}{\consensuscommittee} & 
\cell{\colorscaling}{-} & 
\cell{\colorscaling}{-}  
 \\ \hline

Proof-of-QoS~\cite{yu2019proofofqos} &
\cell{\colorassumption}{\networkpartialsync} & 
\cell{\colorassumption}{\doublecell{\maxfthird}{per committee}} & 
\cell{\colorassumption}{\membershipdynamic} & 
\cell{\colorguarantees}{\safetydeterministic} & 
\cell{\colorguarantees}{\safetydeterministic} & 
\cell{\colorscaling}{\leadersingle} & 
\cell{\colorscaling}{\cryptovrf} & 
\cell{\colorscaling}{-} & 
\cell{\colorscaling}{-} & 
\cell{\colorscaling}{\consensuscommittee} & 
\cell{\colorscaling}{-} & 
\cell{\colorscaling}{-}  
 \\ \hline

Proof-of-Execution~\cite{gupta2021poe} &
\cell{\colorassumption}{\networkpartialsync} & 
\cell{\colorassumption}{\maxfthird} & 
\cell{\colorassumption}{\membershipstatic} & 
\cell{\colorguarantees}{\safetydeterministic} & 
\cell{\colorguarantees}{\safetydeterministic} & 
\cell{\colorscaling}{\leadersingle} & 
\cell{\colorscaling}{\cryptothreshold} & 
\cell{\colorscaling}{\exchangestar} & 
\cell{\colorscaling}{\doublecell{Out-of-order}{processing}} & 
\cell{\colorscaling}{\consensusspeculativeexecution} & 
\cell{\colorscaling}{-} & 
\cell{\colorscaling}{-}  
 \\ \hline

Musch~\cite{jalalzai2018window} &
\cell{\colorassumption}{\networkpartialsync} & 
\cell{\colorassumption}{\maxfthird} & 
\cell{\colorassumption}{\membershipstatic} & 
\cell{\colorguarantees}{\safetydeterministic} & 
\cell{\colorguarantees}{\safetydeterministic} & 
\cell{\colorscaling}{\leadersingle} & 
\cell{\colorscaling}{\cryptomultisignature} & 
\cell{\colorscaling}{\exchangestar} & 
\cell{\colorscaling}{-} & 
\cell{\colorscaling}{-} & 
\cell{\colorscaling}{-} & 
\cell{\colorscaling}{-}  
 \\ \hline

AHL~\cite{dang2019towards} &
\cell{\colorassumption}{\networkpartialsync} & 
\cell{\colorassumption}{\maxfthird} & 
\cell{\colorassumption}{\membershipstatic} & 
\cell{\colorguarantees}{\safetyprobabilistic} & 
\cell{\colorguarantees}{\safetyprobabilistic} & 
\cell{\colorscaling}{\leadersingle} & 
\cell{\colorscaling}{\cryptomultisignature} & 
\cell{\colorscaling}{\exchangestar} & 
\cell{\colorscaling}{-} & 
\cell{\colorscaling}{\consensuscommittee} & 
\cell{\colorscaling}{\parallelizationsharding} & 
\cell{\colorscaling}{HW/TEE}  
 \\ \hline

Dumbo~\cite{guo2020dumbo} &
\cell{\colorassumption}{\networkasync} & 
\cell{\colorassumption}{\maxfthird} & 
\cell{\colorassumption}{\membershipstatic} & 
\cell{\colorguarantees}{\safetyprobabilistic} & 
\cell{\colorguarantees}{\safetyprobabilistic} & 
\cell{\colorscaling}{\leadermulti} & 
\cell{\colorscaling}{\doublecell{\cryptoerasurecoding}{\cryptothresholdencryption}} & 
\cell{\colorscaling}{-} & 
\cell{\colorscaling}{-} & 
\cell{\colorscaling}{\consensuscommittee} & 
\cell{\colorscaling}{-} & 
\cell{\colorscaling}{-}  
 \\ \hline

DP-Hybrid~\cite{zheng2020dphybrid} &
\cell{\colorassumption}{\networkpartialsync} & 
\cell{\colorassumption}{\maxfparameterizable} & 
\cell{\colorassumption}{\membershipdynamic} & 
\cell{\colorguarantees}{\safetyprobabilistic} & 
\cell{\colorguarantees}{\safetyprobabilistic} & 
\cell{\colorscaling}{\doublecell{\leadernone}{\leadersingle}} & 
\cell{\colorscaling}{-} & 
\cell{\colorscaling}{-} & 
\cell{\colorscaling}{-} & 
\cell{\colorscaling}{\consensushierarchical} & 
\cell{\colorscaling}{\parallelizationhierachicalconsensus} & 
\cell{\colorscaling}{-}  
 \\ \hline

DLattice \cite{zhou2019dlattice} &
\cell{\colorassumption}{\networksync} & 
\cell{\colorassumption}{\maxfthird} & 
\cell{\colorassumption}{\membershipdynamic} & 
\cell{\colorguarantees}{\safetyprobabilistic} & 
\cell{\colorguarantees}{\safetyprobabilistic} & 
\cell{\colorscaling}{\leadernone} & 
\cell{\colorscaling}{\cryptovrf} & 
\cell{\colorscaling}{-} & 
\cell{\colorscaling}{-} & 
\cell{\colorscaling}{\consensuscommittee} & 
\cell{\colorscaling}{-} & 
\cell{\colorscaling}{-}  
 \\ \hline

DBFT~\cite{crain2018dbft} &
\cell{\colorassumption}{\networkpartialsync} & 
\cell{\colorassumption}{\maxfthird} & 
\cell{\colorassumption}{\membershipstatic} & 
\cell{\colorguarantees}{\safetydeterministic} & 
\cell{\colorguarantees}{\safetydeterministic} & 
\cell{\colorscaling}{\leadernone} & 
\cell{\colorscaling}{-} & 
\cell{\colorscaling}{-} & 
\cell{\colorscaling}{-} & 
\cell{\colorscaling}{-} & 
\cell{\colorscaling}{-} & 
\cell{\colorscaling}{-}  
 \\ \hline

Cumulus~\cite{gai2021cumulus} &
\cell{\colorassumption}{\networkpartialsync} & 
\cell{\colorassumption}{\maxfthird} & 
\cell{\colorassumption}{\membershipdynamic} & 
\cell{\colorguarantees}{\safetyprobabilistic} & 
\cell{\colorguarantees}{\safetyprobabilistic} & 
\cell{\colorscaling}{\leadersingle} & 
\cell{\colorscaling}{\doublecell{\cryptothreshold}{\cryptovrf}} & 
\cell{\colorscaling}{-} & 
\cell{\colorscaling}{-} & 
\cell{\colorscaling}{-} & 
\cell{\colorscaling}{-} & 
\cell{\colorscaling}{-}  
 \\ \hline

Chainspace~\cite{al-bassam2018chainspace} &
\cell{\colorassumption}{\networkasync} & 
\cell{\colorassumption}{\maxfthird} & 
\cell{\colorassumption}{\membershipdynamic} & 
\cell{\colorguarantees}{\safetyprobabilistic} & 
\cell{\colorguarantees}{\safetyprobabilistic} & 
\cell{\colorscaling}{\leadersingle} & 
\cell{\colorscaling}{-} & 
\cell{\colorscaling}{-} & 
\cell{\colorscaling}{-} & 
\cell{\colorscaling}{-} & 
\cell{\colorscaling}{\parallelizationsharding} & 
\cell{\colorscaling}{-}  
 \\ \hline

CAPER~\cite{amiri2019caper} &
\cell{\colorassumption}{\networkasync} & 
\cell{\colorassumption}{\maxfthird} & 
\cell{\colorassumption}{\membershipstatic} & 
\cell{\colorguarantees}{\safetydeterministic} & 
\cell{\colorguarantees}{\safetydeterministic} & 
\cell{\colorscaling}{\leadersingle} & 
\cell{\colorscaling}{-} & 
\cell{\colorscaling}{-} & 
\cell{\colorscaling}{-} & 
\cell{\colorscaling}{\consensushierarchical} & 
\cell{\colorscaling}{-} & 
\cell{\colorscaling}{-}  
 \\ \hline

BlockTree~\cite{blocktree2021vishwakarma} &
\cell{\colorassumption}{\networkasync} & 
\cell{\colorassumption}{\maxfhalf} & 
\cell{\colorassumption}{\membershipdynamic} & 
\cell{\colorguarantees}{\safetyprobabilistic} & 
\cell{\colorguarantees}{\safetyprobabilistic} & 
\cell{\colorscaling}{\leadersingle} & 
\cell{\colorscaling}{-} & 
\cell{\colorscaling}{-} & 
\cell{\colorscaling}{-} & 
\cell{\colorscaling}{-} & 
\cell{\colorscaling}{\parallelizationsharding} & 
\cell{\colorscaling}{-}  
 \\ \hline

Dispel~\cite{voron2019dispel} &
\cell{\colorassumption}{\networkpartialsync} & 
\cell{\colorassumption}{\maxfthird} & 
\cell{\colorassumption}{\membershipdynamic} & 
\cell{\colorguarantees}{\safetydeterministic} & 
\cell{\colorguarantees}{\safetydeterministic} & 
\cell{\colorscaling}{\leadernone} & 
\cell{\colorscaling}{-} & 
\cell{\colorscaling}{-} & 
\cell{\colorscaling}{\doublecell{\textit{distributed}}{pipeline}} & 
\cell{\colorscaling}{-} & 
\cell{\colorscaling}{-} & 
\cell{\colorscaling}{-}  
 \\ \hline
 
BigBFT \cite{alqahtani2021bigbft} &
\cell{\colorassumption}{\networkpartialsync} & 
\cell{\colorassumption}{\maxfthird} & 
\cell{\colorassumption}{\membershipstatic} & 
\cell{\colorguarantees}{\safetydeterministic} & 
\cell{\colorguarantees}{\safetydeterministic} & 
\cell{\colorscaling}{\leadermulti} & 
\cell{\colorscaling}{\cryptomultisignature} & 
\cell{\colorscaling}{-} & 
\cell{\colorscaling}{\doublecell{pipelining blocks}{across rounds}} & 
\cell{\colorscaling}{-} & 
\cell{\colorscaling}{-} & 
\cell{\colorscaling}{-}  
 \\ \hline

Gosig \cite{li2020gosig} &
\cell{\colorassumption}{\networkpartialsync} & 
\cell{\colorassumption}{\maxfthird} & 
\cell{\colorassumption}{\membershipstatic} & 
\cell{\colorguarantees}{\safetydeterministic} & 
\cell{\colorguarantees}{\safetyprobabilistic} & 
\cell{\colorscaling}{\leaderrandomselection} & 
\cell{\colorscaling}{\cryptomultisignature} & 
\cell{\colorscaling}{\exchangegossip} & 
\cell{\colorscaling}{\doublecell{pipelines gossiplayer}{\textit{and} BFT protocol}} & 
\cell{\colorscaling}{-} & 
\cell{\colorscaling}{-} & 
\cell{\colorscaling}{-}  
 \\ \hline

RingBFT \cite{rahnama2021ringbft} &
\cell{\colorassumption}{\networkpartialsync} & 
\cell{\colorassumption}{\maxfthird} & 
\cell{\colorassumption}{\membershipstatic} & 
\cell{\colorguarantees}{\safetydeterministic} & 
\cell{\colorguarantees}{\safetydeterministic} & 
\cell{\colorscaling}{\leadersingle} & 
\cell{\colorscaling}{-} & 
\cell{\colorscaling}{\exchangering} & 
\cell{\colorscaling}{-} & 
\cell{\colorscaling}{-} & 
\cell{\colorscaling}{\parallelizationsharding} & 
\cell{\colorscaling}{-}  
 \\ \hline

DispersedLedger \cite{yang2022dispersed} &
\cell{\colorassumption}{\networkasync} & 
\cell{\colorassumption}{\maxfthird} & 
\cell{\colorassumption}{\membershipstatic} & 
\cell{\colorguarantees}{\safetydeterministic} & 
\cell{\colorguarantees}{\safetydeterministic} & 
\cell{\colorscaling}{\leadernone} & 
\cell{\colorscaling}{\cryptoerasurecoding} & 
\cell{\colorscaling}{-} & 
\cell{\colorscaling}{-} & 
\cell{\colorscaling}{-} & 
\cell{\colorscaling}{-} & 
\cell{\colorscaling}{-}  
 \\ \hline

Saguaro \cite{amiri2021saguaro} &
\cell{\colorassumption}{\networkpartialsync} & 
\cell{\colorassumption}{\maxfthird} & 
\cell{\colorassumption}{\membershipdynamic} & 
\cell{\colorguarantees}{\safetydeterministic} & 
\cell{\colorguarantees}{\safetydeterministic} & 
\cell{\colorscaling}{\leadersingle} & 
\cell{\colorscaling}{\cryptothreshold} & 
\cell{\colorscaling}{-} & 
\cell{\colorscaling}{-} & 
\cell{\colorscaling}{\consensushierarchical} & 
\cell{\colorscaling}{\parallelizationhierachicalconsensus} & 
\cell{\colorscaling}{-}  
 \\ \hline


Qanaat \cite{amiri2021qanaat} &
\cell{\colorassumption}{\networkpartialsync} & 
\cell{\colorassumption}{\maxfthird} & 
\cell{\colorassumption}{\membershipstatic} & 
\cell{\colorguarantees}{\safetydeterministic} & 
\cell{\colorguarantees}{\safetydeterministic} & 
\cell{\colorscaling}{\leadersingle} & 
\cell{\colorscaling}{\cryptosecretsharing} & 
\cell{\colorscaling}{-} & 
\cell{\colorscaling}{-} & 
\cell{\colorscaling}{-} & 
\cell{\colorscaling}{\parallelizationsharding} & 
\cell{\colorscaling}{-}  
 \\ \hline



Narwhal-HotStuff \cite{danezis2022narwhal} &
\cell{\colorassumption}{\networkpartialsync} & 
\cell{\colorassumption}{\maxfthird} & 
\cell{\colorassumption}{\membershipstatic} & 
\cell{\colorguarantees}{\safetydeterministic} & 
\cell{\colorguarantees}{\safetydeterministic} & 
\cell{\colorscaling}{\leaderrotating} & 
\cell{\colorscaling}{\cryptothreshold} & 
\cell{\colorscaling}{\exchangestar} & 
\cell{\colorscaling}{\pipelininghotstuff} & 
\cell{\colorscaling}{-} & 
\cell{\colorscaling}{-} & 
\cell{\colorscaling}{-}  
 \\ \hline

Tusk \cite{danezis2022narwhal} &
\cell{\colorassumption}{\networkasync} & 
\cell{\colorassumption}{\maxfthird} & 
\cell{\colorassumption}{\membershipstatic} & 
\cell{\colorguarantees}{\safetyprobabilistic} & 
\cell{\colorguarantees}{\safetyprobabilistic} & 
\cell{\colorscaling}{\leaderrotating} & 
\cell{\colorscaling}{\cryptothreshold} & 
\cell{\colorscaling}{\exchangestar} & 
\cell{\colorscaling}{\pipelininghotstuff} & 
\cell{\colorscaling}{-} & 
\cell{\colorscaling}{-} & 
\cell{\colorscaling}{-}  
 \\ \hline

Avalanche \cite{rocket2020avalanche} &
\cell{\colorassumption}{\networkpartialsync} & 
\cell{\colorassumption}{\maxfparameterizable} & 
\cell{\colorassumption}{\membershipdynamic} & 
\cell{\colorguarantees}{\safetyprobabilistic} & 
\cell{\colorguarantees}{\safetyprobabilistic} & 
\cell{\colorscaling}{\leadernone} & 
\cell{\colorscaling}{-} & 
\cell{\colorscaling}{\exchangegossip} & 
\cell{\colorscaling}{-} & 
\cell{\colorscaling}{\consensusrandomizedsampling} & 
\cell{\colorscaling}{-} & 
\cell{\colorscaling}{-}  
 \\ \hline


C-PBFT \cite{xu2021concurrent} &
\cell{\colorassumption}{\networkpartialsync} & 
\cell{\colorassumption}{\maxfthird} & 
\cell{\colorassumption}{\membershipstatic} & 
\cell{\colorguarantees}{\safetydeterministic} & 
\cell{\colorguarantees}{\safetydeterministic} & 
\cell{\colorscaling}{\leadersingle} & 
\cell{\colorscaling}{-} & 
\cell{\colorscaling}{-} & 
\cell{\colorscaling}{-} & 
\cell{\colorscaling}{\consensushierarchical} & 
\cell{\colorscaling}{\parallelizationhierachicalconsensus} & 
\cell{\colorscaling}{-}  
 \\ \hline


DRBFT \cite{zhan2021drbft} &
\cell{\colorassumption}{\networkpartialsync} & 
\cell{\colorassumption}{\maxfthird} & 
\cell{\colorassumption}{\membershipdynamic} & 
\cell{\colorguarantees}{\safetydeterministic} & 
\cell{\colorguarantees}{\safetydeterministic} & 
\cell{\colorscaling}{\leadersingle} & 
\cell{\colorscaling}{-} & 
\cell{\colorscaling}{-} & 
\cell{\colorscaling}{-} & 
\cell{\colorscaling}{\consensuscommittee} & 
\cell{\colorscaling}{-} & 
\cell{\colorscaling}{-}  
 \\ \hline

SBFT \cite{gueta2019sbft} &
\cell{\colorassumption}{\networksync} & 
\cell{\colorassumption}{\maxfSBFT} & 
\cell{\colorassumption}{\membershipstatic} & 
\cell{\colorguarantees}{\safetydeterministic} & 
\cell{\colorguarantees}{\safetydeterministic} & 
\cell{\colorscaling}{\leadersingle} & 
\cell{\colorscaling}{\cryptothreshold} & 
\cell{\colorscaling}{\exchangestar} & 
\cell{\colorscaling}{\doublecell{learning heuristic} {for dynamic batching}} & 
\cell{\colorscaling}{-} & 
\cell{\colorscaling}{-} & 
\cell{\colorscaling}{-}  
 \\ \hline

Cerberus \cite{hellings2020cerberus} &
\cell{\colorassumption}{\networkasync} & 
\cell{\colorassumption}{\maxfthird} & 
\cell{\colorassumption}{\membershipstatic} & 
\cell{\colorguarantees}{\safetydeterministic} & 
\cell{\colorguarantees}{\safetydeterministic} & 
\cell{\colorscaling}{\leadersingle} & 
\cell{\colorscaling}{-} & 
\cell{\colorscaling}{-} & 
\cell{\colorscaling}{-} & 
\cell{\colorscaling}{-} & 
\cell{\colorscaling}{\parallelizationsharding} & 
\cell{\colorscaling}{-}  
 \\ \hline

GearBox \cite{david2021gearbox} &
\cell{\colorassumption}{\networkpartialsync} & 
\cell{\colorassumption}{\maxfthird} & 
\cell{\colorassumption}{\membershipdynamic} & 
\cell{\colorguarantees}{\safetydeterministic} & 
\cell{\colorguarantees}{\safetyprobabilistic} & 
\cell{\colorscaling}{\leadersingle} & 
\cell{\colorscaling}{-} & 
\cell{\colorscaling}{-} & 
\cell{\colorscaling}{-} & 
\cell{\colorscaling}{\consensuscommittee} & 
\cell{\colorscaling}{\parallelizationsharding} & 
\cell{\colorscaling}{-}  
 \\ \hline

ByShard \cite{hellings2021byshard} &
\cell{\colorassumption}{\conf} & 
\cell{\colorassumption}{\maxfthird} & 
\cell{\colorassumption}{\conf} & 
\cell{\colorguarantees}{\conf} & 
\cell{\colorguarantees}{\conf} & 
\cell{\colorscaling}{\leadersingle} & 
\cell{\colorscaling}{-} & 
\cell{\colorscaling}{-} & 
\cell{\colorscaling}{-} & 
\cell{\colorscaling}{\consensushierarchical} & 
\cell{\colorscaling}{\parallelizationsharding} & 
\cell{\colorscaling}{-}  
 \\ \hline

ICC0 / ICC1 \cite{camenisch2022internet} &
\cell{\colorassumption}{\networkpartialsync} & 
\cell{\colorassumption}{\maxfthird} & 
\cell{\colorassumption}{\membershipdynamic} & 
\cell{\colorguarantees}{\safetydeterministic} & 
\cell{\colorguarantees}{\safetydeterministic} & 
\cell{\colorscaling}{\doublecell{\leadersingle}{\leaderrotating}} & 
\cell{\colorscaling}{\cryptothreshold} & 
\cell{\colorscaling}{\exchangegossip} & 
\cell{\colorscaling}{-} & 
\cell{\colorscaling}{-} & 
\cell{\colorscaling}{-} & 
\cell{\colorscaling}{-}  
 \\ \hline

ICC2 \cite{camenisch2022internet} &
\cell{\colorassumption}{\networkpartialsync} & 
\cell{\colorassumption}{\maxfthird} & 
\cell{\colorassumption}{\membershipdynamic} & 
\cell{\colorguarantees}{\safetydeterministic} & 
\cell{\colorguarantees}{\safetydeterministic} & 
\cell{\colorscaling}{\doublecell{\leadersingle}{\leaderrotating}} & 
\cell{\colorscaling}{\doublecell{\cryptothreshold}{\cryptoerasurecoding}} & 
\cell{\colorscaling}{\exchangegossip} & 
\cell{\colorscaling}{-} & 
\cell{\colorscaling}{-} & 
\cell{\colorscaling}{-} & 
\cell{\colorscaling}{-}  
 \\ \hline

Jolteon \cite{gelashvili2021jolteon} &
\cell{\colorassumption}{{\networkpartialsync}} & 
\cell{\colorassumption}{\maxfthird} & 
\cell{\colorassumption}{\membershipstatic} & 
\cell{\colorguarantees}{{\safetydeterministic}} & 
\cell{\colorguarantees}{\safetydeterministic} & 
\cell{\colorscaling}{\leadersingle} & 
\cell{\colorscaling}{\cryptothreshold} & 
\cell{\colorscaling}{\exchangestar} & 
\cell{\colorscaling}{\pipelininghotstuff} & 
\cell{\colorscaling}{-} & 
\cell{\colorscaling}{-} & 
\cell{\colorscaling}{-}  
 \\ \hline

Ditto \cite{gelashvili2021jolteon} &
\cell{\colorassumption}{{\networkasync}} & 
\cell{\colorassumption}{\maxfthird} & 
\cell{\colorassumption}{\membershipstatic} & 
\cell{\colorguarantees}{{\safetyprobabilistic}} & 
\cell{\colorguarantees}{\safetyprobabilistic} & 
\cell{\colorscaling}{\leadersingle} & 
\cell{\colorscaling}{\cryptothreshold} & 
\cell{\colorscaling}{\exchangestar} & 
\cell{\colorscaling}{\pipelininghotstuff} & 
\cell{\colorscaling}{-} & 
\cell{\colorscaling}{-} & 
\cell{\colorscaling}{-}  
 \\ \hline


Algorand \cite{gilad2017algorand} &
\cell{\colorassumption}{\networksync} & 
\cell{\colorassumption}{\maxfthird} & 
\cell{\colorassumption}{\membershipdynamic} & 
\cell{\colorguarantees}{\safetyprobabilistic} & 
\cell{\colorguarantees}{\safetyprobabilistic} & 
\cell{\colorscaling}{\leaderrandomselection} & 
\cell{\colorscaling}{\cryptovrf} & 
\cell{\colorscaling}{\exchangegossip} & 
\cell{\colorscaling}{-} & 
\cell{\colorscaling}{\consensuscommittee} & 
\cell{\colorscaling}{-} & 
\cell{\colorscaling}{-} 
\\ \hline

\end{tabular}%
}
\end{table*}

\subsection{Summary of BFT Protocols}
All discussed protocols are shown in \cref{tab:protocols}.
We list their assumptions regarding the synchrony model, the number of faults that can be tolerated, membership of nodes (\ie whether nodes are static or can dynamically join or leave the network), and the guarantees for safety and liveness.
Further, we give an overview of which scaling techniques have been employed and combined in the protocols.
PBFT, the starting point for many protocols, assumes a partially synchronous network of static nodes, whereas many of the blockchain BFT protocols target a dynamic network of nodes or the asynchronous model.
Many protocols replace PBFT's single leader with multiple leaders to share the load and reduce its bottleneck, or with a leaderless approach.
Not all blockchain BFT protocols keep PBFT's clique communication; we take clique as the default and list in \cref{tab:protocols} which protocols deviate from this pattern.
While many protocols target improving scalability within the consensus group while increasing the number of nodes, protocols tagged as \enquote{committee}, \enquote{sharding}, and \enquote{hierarchical} generally target scalability and performance improvements by using subsets of nodes participating in consensus.
Some listed approaches are generic frameworks that are not fixed to one specific protocol, \eg RCC~\cite{rcc2021gupta}, Ostraka~\cite{manuskin2020ostraka}, Mitosis~\cite{marson2021mitosis}, RingBFT~\cite{rahnama2021ringbft}, and ByShard~\cite{hellings2021byshard}.
Here, we list assumptions and guarantees if explicitly stated in the corresponding papers or list them as configurable if applicable.

\section{Related Work}
\label{sect:related_work}
Several surveys systematically analyze blockchain protocols; some contain a more or less detailed treatment of scalability. 

Alsunaidi et al. created a survey of blockchain consensus algorithms, focusing on performance and security~\cite{Alsunaidi19Consensus}. The authors distinguish between \textit{proof-based} (\eg PoW) and \textit{voting-based} (\eg BFT). Scalability is a mentioned challenge, but is not analyzed beyond categorizing proof-based protocols as \enquote{strongly} and vote-based protocols as \enquote{weakly} scalable.

Bano et al. presented an SoK about blockchain consensus protocols~\cite{bano19sok}, where the main contribution is a systematization framework that tracks the chronological evolution of blockchain consensus protocols and a categorization using this framework. While the framework explains that consensus scalability can be achieved by advancing from \textit{hybrid single committee consensus} to \textit{hybrid multiple committee consensus} (\eg through sharding), it does not treat scalability mechanics for consensus in general, \eg for consensus within a single committee.

Berger et al. created a short survey in 2018 that broadly analyzes scalability techniques used in BFT consensus protocols~\cite{berger18scaling}. Possibly, our survey can be best understood as progressing this effort by (1) using a systematic methodology, (2) increasing the level of detail, and (3) applying the analysis to contemporary research works, thus extending the scope by many papers that have been published just recently.


Ferdous et al.'s survey on block\-chain consensus introduced a taxonomy of desirable properties~\cite{ferdous2020blockchain}. While scalability is one such property, the technical aspects are not discussed.

Hafid et al. analyze the scalability of blockchain platforms with a focus on first and second-layer solutions~\cite{hafid2020scaling}, \ie changes to the blockchain, \eg the block structure using DAGs or increasing block sizes, and mechanisms implemented outside of it, \eg side-chains, child-chains, or payment channels.
The authors propose a taxonomy based on committee formation and consensus within a committee and compare sharding-based protocols.
Scalability outside of sharding is not considered.

Huang et al. analyze blockchain surveys and focus on theoretical modeling, analysis models, performance measurements, and experiment tools~\cite{huang2021survey}.
Scalability is only considered with regard to sharding and multi-chain interoperability.

Jennath et al. give a general overview of common blockchain consensus protocols, such as \ac{PoW}, \ac{PoS}, \ac{PoET}, \ac{BFT}, and Federated Byzantine agreements~\cite{jennath2020survey}.
%
Lao et al. consider IoT blockchains and their consensus strategies~\cite{lao2020survey}, for which they compare consensus protocols using similar categories. 
Neither survey considers BFT scalability.

Liu et al. analyze recent blockchain techniques and claim that consensus-based scaling is limited, especially with Moore's law nearing its end~\cite{liu2020effective}.
They discuss and evaluate scaling concerning topology and hardware assistance, \eg off-chain or parallel-chain computations or sharding.

Meneghetti et al. presented a survey on blockchain scalability~\cite{meneghetti2019survey}; however, they focus more on smart contract executions, particularly sharding, than on the consensus mechanism.

Monrat et al.'s survey on blockchain applications, challenges, and opportunities~\cite{monrat2019survey} provides a blockchain taxonomy and describes potential applications.
The authors also describe common consensus algorithms but do not focus on scalability. 

Salimitari and Chatterjee created a survey on blockchain consensus protocols in IoT~\cite{salimitari2018survey}. They evaluate various blockchain consensus protocols for use in IoT scenarios. The survey categorizes consensus protocols in rough categories such as \ac{PoW}, \ac{PoS}, \ac{BFT}, VRF-based, and sharding-based solutions.

Vukolić contrasts \ac{PoW}-based algorithms to BFT SMR protocols~\cite{vukolic2015quest}. Vukolić identifies scalability to many consensus nodes as a blocker for the adoption of blockchain consensus. As of 2016, Vukolić identifies optimistic BFT protocols and relaxed fault models such as XFT or hybrid fault models with trusted hardware as potential solutions.
\section{Conclusions and Open Challenges}
\label{sect:conclusions}

One of the major ongoing challenges in the field of blockchain is making Byzantine consensus applicable to large-scale environments. To address this challenge, a large body of research has focused on developing novel techniques to improve the scalability of BFT consensus, paving the way for a new generation of BFT protocols tailored to the needs of blockchain. 
In this SoK paper, we employed a systematic literature search to explore the design space of recent BFT protocols along with their ideas for scaling up to hundreds or thousands of nodes. 
We created a taxonomy of scalability-enhancing techniques, which categorizes these ideas into communication and coordination strategies, pipelining, cryptographic primitives, independent groups, committee selection, and trusted hardware support. 
As shown in \cref{tab:protocols}, many BFT protocols employ not only one idea but rather a combination of several ideas. We also see that a less vigorously explored research field seems to be the incorporation of trusted execution environments, which is inviting for future research works.
Further, we comprehensively discussed all ideas on an abstract level and pinpointed the design space from which their corresponding BFT protocols originated.

Some open challenges regarding BFT scalability remain:
%
%
While we have identified and categorized these scalability techniques, we cannot compare their \emph{effectiveness} solely on the basis of the papers' evaluation results, and a common evaluation platform for these protocols is not yet available~\cite{amiri2022bedrock}.
\cref{tab:protocols} shows a wide range of combinations of techniques that have so far been applied; however, \emph{further combinations} may exist that lead to valid, performant, and highly scalable protocols and which have to be identified.
Not only the techniques' effectiveness is important, but their \emph{complexity} regarding computational resource requirements, implementation effort, or proof of correctness in a protocol is also relevant as well and may differ widely.
Finally, depending on the specific \textit{application requirements}, different agreement protocols may be more suitable for specific deployment settings than others. 
As the BFT protocol landscape is extensive, developing a guideline for selecting the most fitting protocol according to these requirements may be helpful.

\section*{Acknowledgments}

This work has been funded by the Deutsche Forschungsgemeinschaft (DFG, German Research Foundation) grant number 446811880 (BFT2Chain).

\balance
\bibliographystyle{IEEEtran}
\bibliography{main.bib}

\end{document}